\documentclass[aps,prx,onecolumn,amssymb,amsfonts]{revtex4-2}  

\usepackage{natbib}
\usepackage{graphicx}
\usepackage{amsmath}
\usepackage[inline]{enumitem}
\usepackage{textcomp} 
\usepackage[mathscr]{euscript}
\usepackage{bbm}
\usepackage{setspace} 
\usepackage{xcolor}

\usepackage{xr}
\newcommand{\alex}[1]{\textcolor{black}{#1}}
\newcommand{\alexx}[1]{\textcolor{black}{#1}}
\newcommand{\al}[1]{\textcolor{black}{#1}}
\newcommand{\all}[1]{\textcolor{black}{#1}}
\newcommand{\rev}[1]{\textcolor{black}{#1}}
\begin{document}

\title{Manifestation of aberrations in full-field optical coherence tomography}

\author{Victor Barolle,$^\dag$ Jules Scholler,$^\dag$ Pedro Mec\^{e}, Jean-Marie Chassot, Kassandra Groux, Mathias Fink, A. Claude Boccara, and Alexandre Aubry*}

\affiliation{Institut Langevin, ESPCI Paris, CNRS, PSL University, 1 rue Jussieu, 75005 Paris, France\\
$^\dag$These authors equally contributed to this work \\
*Corresponding author: alexandre.aubry@espci.fr}





\begin{abstract}
We report on a theoretical model for image formation in full-field optical coherence tomography (FFOCT). Because the spatial incoherence of the illumination acts as a virtual confocal pinhole in FFOCT, its imaging performance is equivalent to a scanning time-gated coherent confocal microscope. In agreement with optical experiments enabling a precise control of aberrations, FFOCT is shown to have nearly twice the resolution of standard imaging at moderate aberration level. Beyond a rigorous study on the sensitivity of FFOCT with respect to aberrations, this theoretical model paves the way towards an optimized design of adaptive optics and \rev{computational tools} for high-resolution and deep imaging of biological tissues.
\end{abstract}

\maketitle

\section{Introduction}
\label{sec1}

Since the early nineties, optical coherence tomography (OCT) has become a widely used imaging technique in medical science especially in ophthalmology. The use of a broadband light source coupled with an interferometric device has allowed non-invasive imaging with a micrometric resolution up to one millimeter deep \alex{\cite{Dunsby,badon2017multiple}}. From a certain point of view, OCT can be considered as an optical analog of ultrasound imaging where the reflectivity of an object is obtained by measuring the reflected echoes of the medium to image. As there is no instrument fast enough to measure the amplitude and phase of the reflected field, OCT relies on phase-shifting interferometry to do so \cite{wang2007vivo}. Therefore, \al{at low numerical aperture,} the axial resolution \al{is dictated by} the light source spectrum (LED, halogen light, femtosecond laser, etc.), typically from 1 to 10 $\mu$m, whereas the transverse resolution \al{only} depends on the numerical aperture of the microscope objectives. 

The original Time-Domain OCT (TD-OCT) needs a raster scanning of the sample \cite{Huang_91}. The acquisition time is thus particularly long as it scales as the number of voxels in the 3D image of the sample. In the late nineties, two main breakthroughs appeared with Spectral-Domain OCT (SD-OCT) and FFOCT. On one hand, SD-OCT relies on the spectral decomposition of the reflected light using a spectrometer \cite{Choma_03, Yun_03, deBoer_03}. It is then possible to directly measure \alex{reflectivity axial profiles (A-scans or A-lines)}, thereby drastically decreasing the acquisition time of the image. On the other hand, FFOCT uses a spatially incoherent light source illuminating a sample over the whole field of view \cite{Beaurepaire_98, Dubois_04}. In FFOCT, $\textit{en face}$ images are directly measured in parallel thanks to a bi-dimensional detector (CCD or CMOS camera), hence without \alex{any mirror galvanometer}. \al{Originally, FFOCT has been developed in the time domain. An axial translation of the sample should then be performed to obtain a volumetric image of the sample. Interestingly, FFOCT can also be performed in the Fourier domain using a swept laser source~\cite{Povazay2006}. Working in the Fourier domain implies a much better sensitivity since the signal-to-noise ratio scales as the number of independent wavelengths over which the FFOCT signal is recorded~\cite{Leitgeb2003,Boer2003,Choma2003}. The use of high-speed cameras can then lead to real time volumetric imaging~\cite{Hillmann2016,Hillmann_2016}. Structured~\cite{Grebenyuk2018} or spatially-incoherent~\cite{Auksorius2020} illuminations can also be used to maintain the confocal gate of the original FFOCT scheme~\cite{Beaurepaire_98, Dubois_04} and filter most of multiple scattering background.} 

\al{Whether it be recorded in the time or spectral domains,} studying the impact of aberrations on the \al{FFOCT} image formation is crucial to understand how to incorporate efficient adaptive optics~\cite{Xiao_16_jbo,Scholler:20} or post-processing \al{computational} tools \rev{such as interferometric synthetic aperture microscopy~\cite{Ralston2007,Adie2012,Ahmad2013,Hillmann2016} or matrix imaging methods~\cite{Kang2017,Badon2019}} in FFOCT~\cite{Barolle2019}. Indeed, system-induced and sample-induced aberrations are the main fundamental limits \all{in the quest for a diffraction-limited resolution,  a high sensitivity and a large penetration  depth.} Interestingly, \rev{previous} studies \rev{showed} that FFOCT is extremely robust to defocus \rev{under spatially-incoherent illumination} \cite{Xiao_16_osa, Xiao_16_jbo}. \al{Furthermore, despite of ocular aberrations~\cite{jarosz2017high}, FFOCT successfully achieved cellular resolution for in-vivo human retinal photoreceptor imaging {close to the foveal center, where confocal scanning laser ophthalmoscopy} failed \cite{mece2020coherence}. \rev{On the contrary, a recent experimental study showed that FFOCT is clearly sensitive to aberrations~\cite{Blavier2021}. }} These \rev{striking but seemingly contradictory} results motivated the present work.

After reminding briefly the principle of \al{time domain} FFOCT, we first revisit those previous experimental observations \cite{Xiao_16_osa, Xiao_16_jbo} by pointing out an apparent inconsistency. While FFOCT seems pretty insensitive to out-of-focus for a resolution target, its \alex{imaging point spread function (point-like scatterer)} does not exhibit any clear improvement \alex{under a large defocus} compared to standard incoherent microscopy. To explain this apparent contradiction, an analytical expression of the FFOCT \al{signal} is derived using Fourier optics. \al{\rev{In agreement with previous works~\cite{Marks2009,Tricoli2019} but in a simpler way}, the proposed model accounts for the imaging performance of both time-domain and spectral-domain FFOCT \rev{under a partially coherent illumination}.} FFOCT is shown to be equivalent to a time-gated confocal microscope \all{where the scanner acts synchronously on the sample arm and on the reference arm (at the difference of standard OCT systems where only the object beam is scanned).} \rev{The size of the virtual confocal pinhole is governed by the coherence length of the incident wave-field.} In quantitative agreement with the experimental results, the imaging performance of FFOCT is shown to strongly depend on the nature of the object and of aberrations:
\begin{itemize}
\item (\textit{i}) For a coherent object, FFOCT is shown to be particularly robust to low-order symmetric aberrations (such as defocus) compared to incoherent imaging. The optical transfer function in incoherent imaging is given by the autocorrelation function of the pupil function. In coherent imaging (FFOCT), this autocorrelation function shall be weighted by the object's spectrum. Therefore, if the object's spectrum and the aberration phase shift are, respectively, decreasing and increasing functions with respect to spatial frequency, then the resolution and contrast of the FFOCT image are clearly improved compared to a standard \al{incoherent} image. 
\item (\textit{ii}) On the contrary, anti-symmetric aberrations (such as coma) manifest themselves as a spatial frequency filter. In that singular case, FFOCT yields an \al{incoherent-like} confocal image. 
\item (\textit{iii}) For \al{random} media or point-like objects, the equivalence with a \all{time gated} confocal microscope holds whatever the nature of aberrations. \all{For a defocus smaller than the depth-of-field, FFOCT resolution is improved by a factor two compared to standard incoherent imaging. For a larger defocus, FFOCT resolution is dictated by the geometrical optics limit.}
\item (\textit{iv}) For sample-induced aberrations, a random Gaussian phase screen~\cite{mertz2015field} can be used as a \alex{preliminary} model. The corresponding PSF is then the sum of diffraction-limited (ballistic) and aberrated (scattering) components. Compared to a standard incoherent microscope, the FFOCT PSF is shown to exhibit an enhancement by a factor 2 of the ballistic component in a strong aberration regime. The spatial extension of the aberrated component is also reduced by a factor two, which is again a manifestation of the confocal feature of FFOCT.
\end{itemize}
After the demonstration of these main results, a discussion will be made about the limits of our current model, in particular the impact of time gating, the deformation of the coherence plane as well as the multiple scattering contribution that will unavoidably predominate at large depths. We will conclude on how adaptive optics and matrix imaging methods can help, in the near future, on how to overcome the aforementioned issues. 

\section{Experimental results}

\begin{figure}[htbp]
\hspace*{-0.3cm}
\centering
\includegraphics[width=8.5cm]{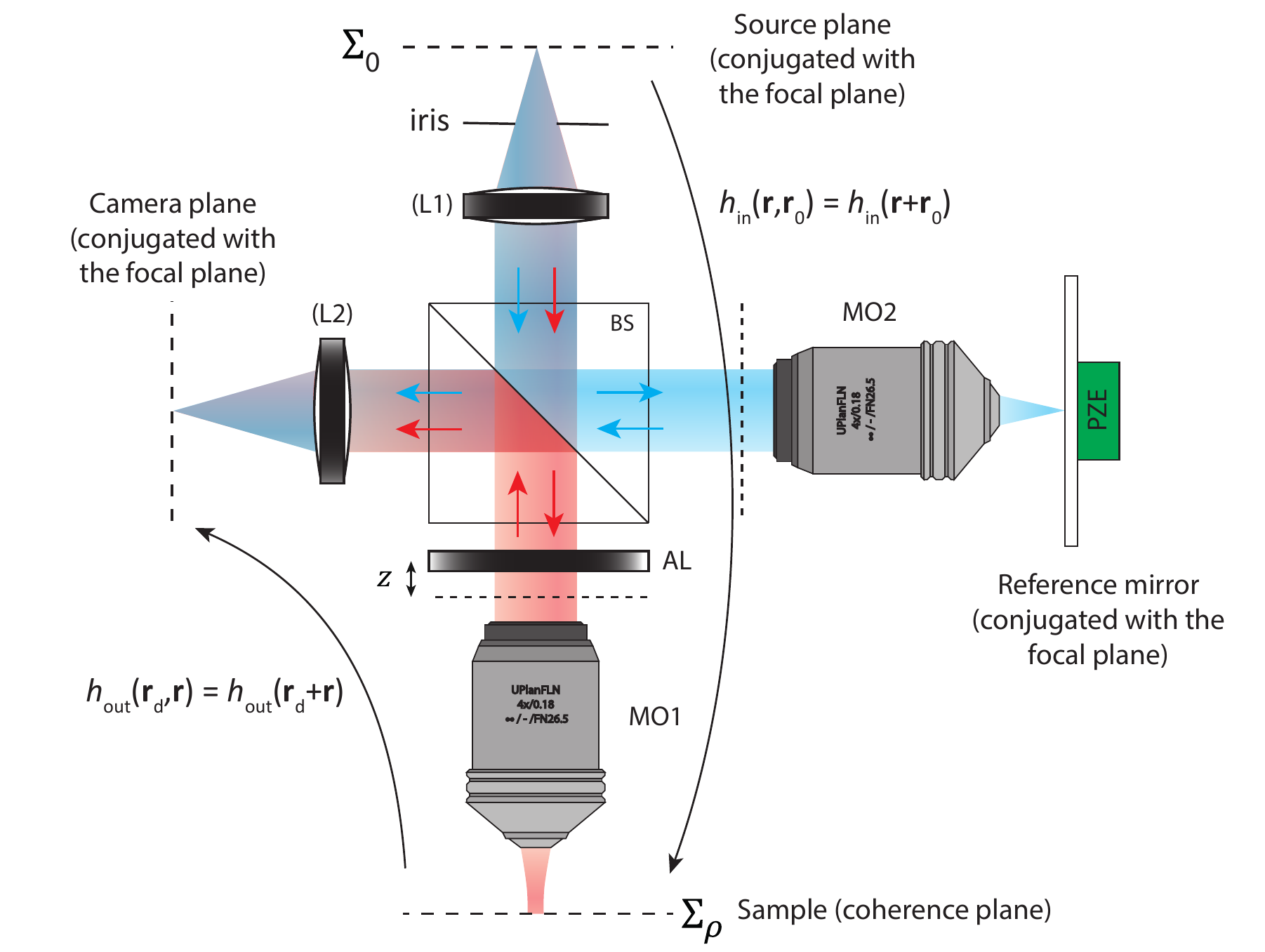}
\caption{FFOCT set up. (L1), (L2) thin lenses. (BS) beamsplitter. (MO1), (MO2) microscope objectives. (PZE) piezoelectric actuator. (AL) Adaptive Lens.}
\label{fig1}
\end{figure}

The principle of \alex{time domain} full-field OCT relies on low-coherence interference microscopy. The experimental set up is based on a Michelson interferometer with identical microscope objectives \alex{(MO)} in both arms, as depicted in Fig.~\ref{fig1}(a). This configuration is referred to as the Linnik interferometer. In the first arm, a reference mirror is placed in the focal plane of the microscope objective. The second arm contains the scattering sample to be imaged. The same broadband incoherent light source is used to illuminate the entire field of the microscope objectives. \rev{To obtain a sufficient concentration and an homogeneous distribution of incident light over the field-of-wiew, a critical or Köhler illumination scheme can be used~\cite{Born}. As illustrated in Fig.~\ref{fig1}, an iris is placed in front of the source. It limits the numerical aperture $\sin \alpha_\textrm{in}$ of the imaging device at input}. Because of the broad spectrum of \rev{the incident} light, some interferences occur between the two arms provided that the \alex{optical} path difference through the interferometer is close to zero. The length of the reference arm determines the slice of the sample to be imaged. The backscattered light from each voxel of this slice can only interfere with the light coming from the conjugated point of a reference mirror. The spatial incoherence of the light source indeed acts as a physical pinhole. All these interference signals are recorded in parallel by the pixels of \alex{a CMOS or CCD camera} \alex{in} the imaging plane. The interference term is extracted from the recorded intensity by phase shifting interferometry (“four phases method”~\cite{Dubois_04}) using a piezoelectric actuator placed on the reference mirror.

\begin{figure*}[htbp]
\centering
\includegraphics[width=\linewidth]{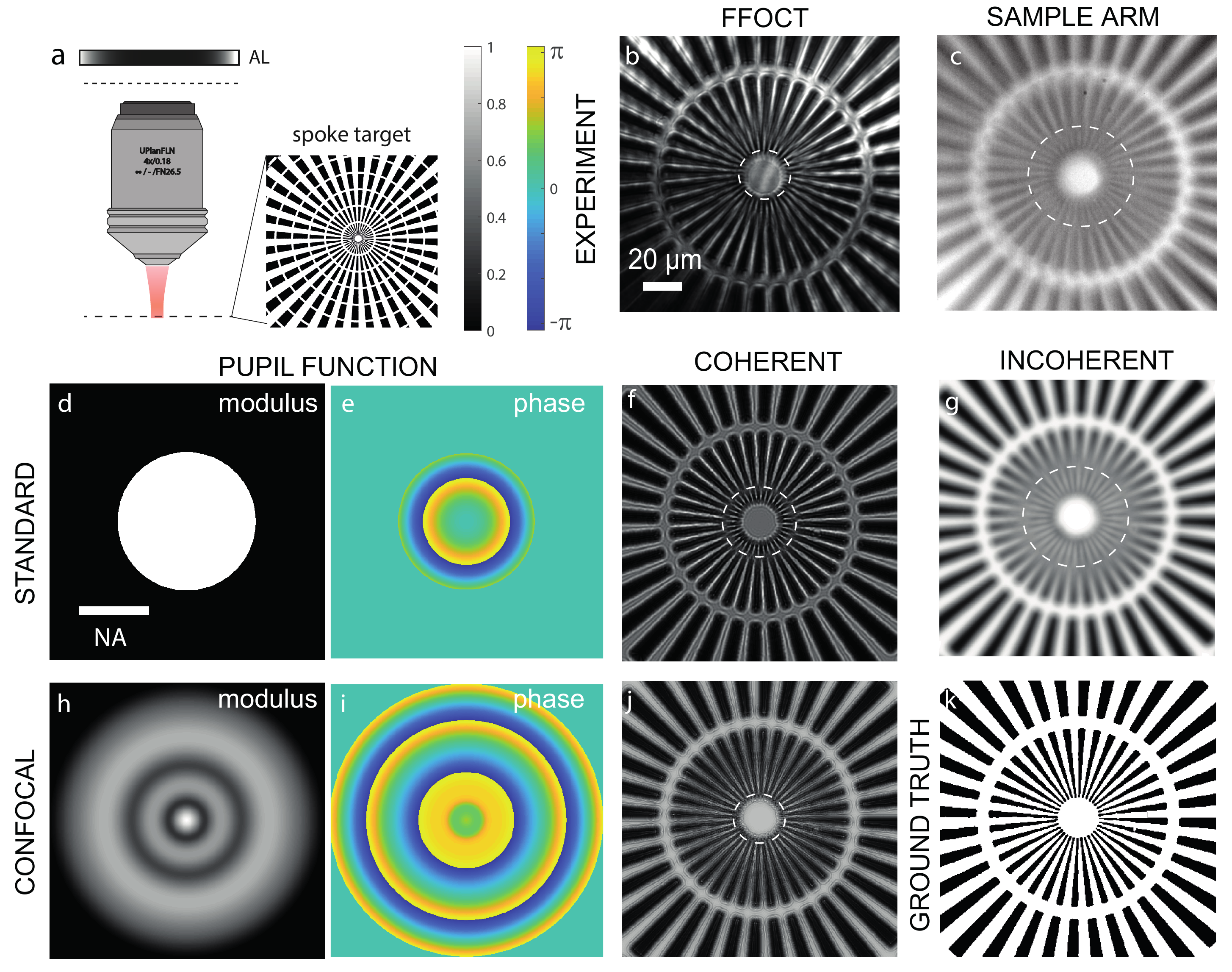}
\caption{Imaging of a Siemens target under a defocus \all{$z=3.7$} $\mu$m \all{that corresponds to a focal spot radius $\delta_\infty \sim 3.6$ $\mu$m (Eq.~\ref{dgeom}).} \al{(a) Experimental configuration}. \al{(b) FFOCT image \rev{under a spatially-incoherent illumination}. (c) Standard incoherent image obtained by blocking the reference arm in Fig.~\ref{fig1}.} (d,e) Modulus and phase of the pupil function $\mathcal{H}_\textrm{out}$. \rev{(f) Theoretical prediction of the FFOCT image under a coherent illumination (Eq.~\ref{terme_interference3} with $h_2=h_{\textrm{out}}$). (g) Theoretical prediction of the incoherent image (Eq.\ref{image_conventionnelle}).} (h,i) Modulus and phase of the confocal pupil function $\mathcal{H}_2$. \al{(j) Theoretical prediction for the confocal coherent image (Eq. ~\ref{terme_interference3} \rev{with $h_2=h_{\textrm{out}}^2$}). (k) Ground \all{truth} reflectivity of the spoke target.} The modulus of the pupil functions and the image intensities have been normalized by their maximum. \al{The white dashed line circle on each target image represents the spatial frequency cutoff at which the first contrast inversion takes place for the corresponding imaging method.}}
\label{fig2}
\end{figure*}
To investigate the impact of aberrations on FFOCT, an adaptive lens \alex{(AL in Fig.~\ref{fig1})} is introduced in the sample arm [see Fig.~\ref{fig1}]. A wave-front sensor \all{is used to calibrate} the phase distortions induced by the adaptive lens. \alex{A detailed description of the experimental set up is provided in Supplement 1.} Key experimental parameters are the wavelength of the light source (\all{LED, $\lambda \sim 800-900$} nm) as well as the \alex{numerical aperture $\sin \alpha$ of the microscope objectives \all{(Olympus UCPLFLN20X, \rev{$\sin \alpha=0.7$}).} $\alpha$ is the maximal half-angle of the cone of light that can enter or exit the microscope objective.} \rev{The light source fully illuminates the pupil of the microscope objectives, thereby providing a spatially-incoherent illumination of the sample ($\sin \alpha_{\textrm{in}}=\sin \alpha$).} Figures~\ref{fig2}(d) and (e) display the pupil function when a \all{2.5} $\mu$m-defocus is applied by means of the adaptive lens. 
For this first experiment, a Siemens star target is placed in the sample focal plane [\al{Fig.~\ref{fig2}(a)}]. This object is indeed the ideal target for quantifying the optical resolution in coherence microscopy \cite{Coherent_resolution_16} since it gathers the maximum number of spatial frequency in a single image. Figure~\ref{fig2}\al{(c)} shows the direct image of the target obtained by blocking the reference arm. Not surprisingly, due to defocus, this standard incoherent image suffers from an important loss of resolution compared to the initial object [Fig.~\ref{fig2}\al{(k)}]. On the contrary, the FFOCT image, displayed in Fig.~\ref{fig2}\al{(b)}, exhibits a much better resolution. \alex{Unlike a standard incoherent image, the smallest details of the resolution target are revealed by the FFOCT system. Moreover, the first contrast inversion~\cite{Goodman} in the FFOCT image (white dashed circle in Fig.~\ref{fig2}) occurs at a larger spatial frequency cutoff than in the incoherent image. These observations seem} to indicate a low sensitivity of FFOCT with respect to defocus, which is in good agreement with previous experimental observations~\cite{Xiao_16_osa,Xiao_16_jbo, Scholler:20}. 

\begin{figure*}
    \centering
    \includegraphics[width=\linewidth]{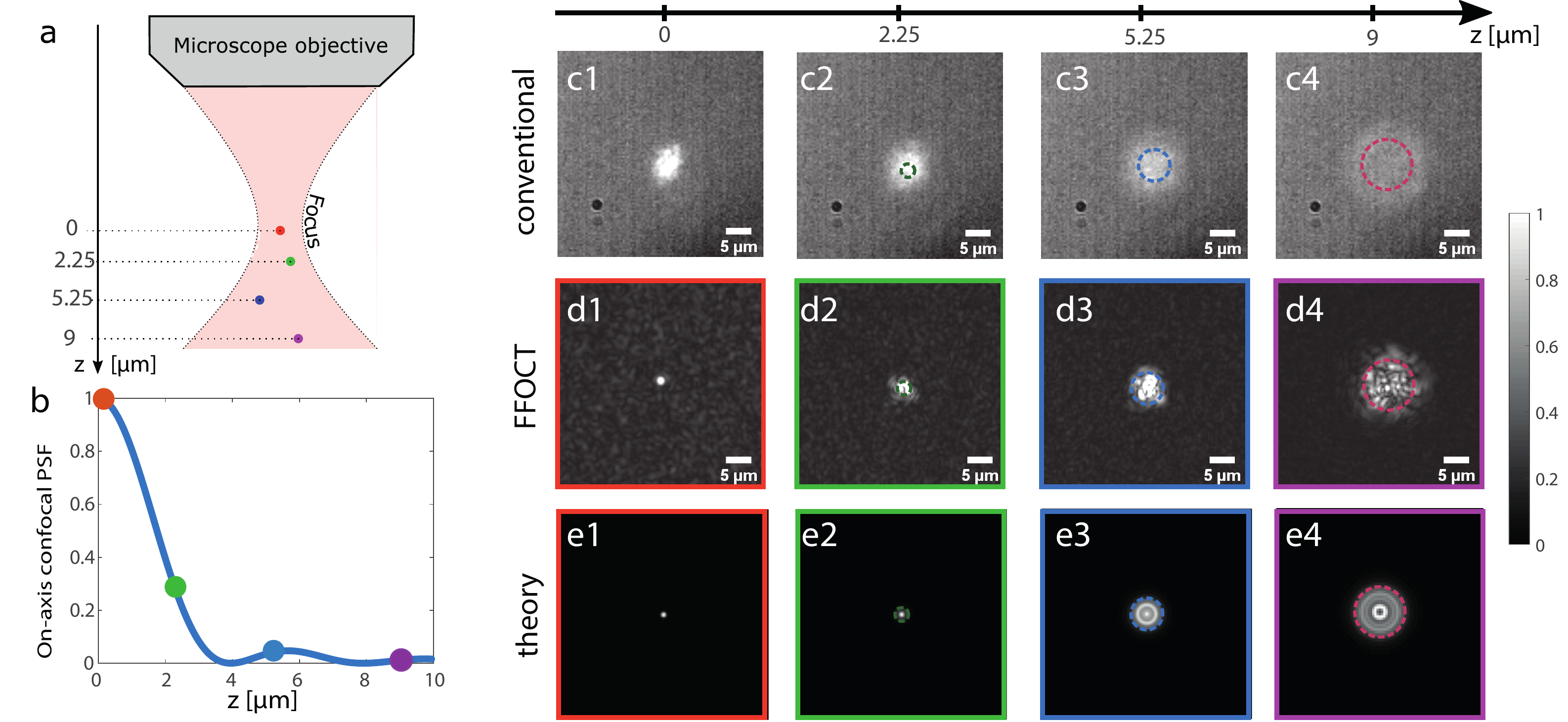}
    \caption{Imaging of \al{micro-beads} for various amounts of defocus.  (a) Scheme of the experimental configuration: The coherence plane is shifted by translating the whole \rev{reference} arm and each defocus distance corresponds to a different nanobead. (b) On-axis evolution of the confocal PSF $h_2(\mathbf{r}=\mathbf{0},z)$ (Eq.~\ref{onaxisPSF}). \rev{(c,d)} Imaging PSF recorded for different amount of defocus in conventional microscopy and FFOCT, respectively. \alex{(e)} Imaging PSF predicted theoretically in FFOCT for the same amount of defocus. The spatial extension \all{$\delta_\infty$} of the geometrical PSF (Eq.~\ref{dgeom}) is shown as a circle in (c,d,e). All the images displayed are normalized by their maximum.}
    \label{fig3}
\end{figure*}

To confirm this experimental observation, a second experiment has been performed with the same set up to extract the FFOCT PSF, without the adaptive lens, but still in presence of defocus [see Fig.~\ref{fig3}(a)]. To that aim, a 3D phantom sample containing 0.05\% (w/v) TiO$_2$ nanobeads with 2\% (w/v) agarose is placed in the field-of-view of the microscope objective. Such a concentration makes the nanobeads isolated from each other. This sparse configuration enables the measurement of the imaging PSF. To do so, the coherence plane and the focal plane are merged such that a single nanobead in the focal plane yields the imaging PSF at focus. As expected, the PSF in FFOCT is only limited by diffraction [Left of Fig.~\ref{fig3}(d1)]. This is in contrast with the incoherent PSF obtained by blocking the reference arm and that already exhibits a blurry feature [Left of Fig.~\ref{fig3}(c1)]. This difference is explained by the agarose sample which is diffusive and induces multiple scattering events upstream of the focal plane. In FFOCT, most of this diffuse contribution is eliminated by the coherent time gating of singly-scattered photons provided by low coherence interferometry. 

In a second step, the reference stage is axially shifted such that the coherence volume now explores out-of-focus \al{micro-beads}. The result is displayed for three different beads located at different de-focus distances inside the Agar gel sample (refractive index $n=1.33$): $z=0$, 2.25, 5.25 and 9 $\mu$m. The incoherent and FFOCT PSFs are displayed in Figs.~\ref{fig3}(c) and (d), respectively. In agreement with the previous experiment depicted in Fig.~\ref{fig2}, FFOCT shows a better robustness to aberrations for a small defocus. However, \all{beyond the depth-of-field ($z\sim 4$ $\mu$m, here)}, the PSF in FFOCT starts to exhibit a speckle-like feature. \alex{This is a manifestation of a deformation of the coherence plane due to the scattering events induced by the agarose gel ahead of the focal plane}. Even worse, its spatial extension increases and becomes comparable with the incoherent PSF. To explain this different behavior of FFOCT at small and large defocus, a theoretical model is needed. Based on Fourier optics, this model will enable a prediction of the FFOCT image whatever the nature of the object and the form of aberrations. 

\section{Analytical model}

In this section, we present a Fourier optics model of the optical imaging process in reflection microscopy under \rev{a partially incoherent illumination}. In particular, analytical expressions for the standard incoherent and FFOCT images are derived. FFOCT is shown to be equivalent to \all{the scanning \rev{time-gated coherent confocal} microscope discussed in Sec.~\ref{sec1}}.

For the sake of simplicity, a scalar model is considered and multiple scattering is neglected. 
\alex{As we will see later, the broad bandwidth $\Delta \omega$ of the light source allows in FFOCT the coherent time gating of photons associated with single scattering events taking place in the coherence plane. In the following, we thus restrict our study to a planar object that matches with the coherence plane. This hypothesis remains valid for the incoherent images displayed in Figs.~\ref{fig2} and ~\al{\ref{fig3}} since the objects considered in this paper are either planar or point-like, respectively.}

\subsection{Incident wave-field}

The propagation between the source plane $\Sigma_{0}$ and the \alex{object} plane $\Sigma_{\rho}$ is described by the impulse response $h_\textrm{in}(\textbf{r},\textbf{r}_0)$ between a point source in \alex{$\Sigma_{0}$} at coordinate $\textbf{r}_0$ and a point in the sample plane at coordinate $\textbf{r}$ (Fig. $\ref{fig1}$). \alex{In the following, the dependence of $h_\textrm{in}(\textbf{r},\textbf{r}_0)$ on frequency $\omega$ is neglected \rev{because of the relatively narrow bandwidth of the light source and the use of \all{achromatic} lenses}. This impulse response can be used to connect the incident electric field in the source and sample planes, \alex{$ E_{0}(\textbf{r}_0,\omega)$ and $ E_{i}(\textbf{r},\omega)$}, respectively: } 
\begin{equation}
    E_{i}(\textbf{r}\alex{,\omega}) = \int_{\Sigma_{0}} h_\textrm{in}(\textbf{r},\textbf{r}_0) E_{0}(\textbf{r}_0\alex{,\omega}) d\textbf{r}_0.
    \label{champ_incident}
\end{equation}
\rev{As we will see, the input impulse $h_\textrm{in}(\textbf{r},\textbf{r}_0)$ can account for: (\textit{i}) the aberration undergone by the incident light $E_i$ from the source plane to the sample plane; (\textit{ii}) its partial coherence that can be controlled by the iris displayed in Fig.~\ref{fig1}.}

\subsection{Reflected wave-field in the sample arm}

Let $\rho(\textbf{r})$ be the sample reflectivity in the coherence plane. Under the single scattering assumption, the reflected wave-field $E_{r}$ can be written as follows in the sample plane:
\begin{equation}
    E_{r}(\textbf{r}\alex{,\omega}) = \rho(\textbf{r}) E_{i}(\textbf{r}\alex{,\omega})= \rho(\textbf{r}) \int_{\Sigma_{0}} h_\textrm{in}(\textbf{r},\textbf{r}_0) E_{0}(\textbf{r}_0\alex{,\omega}) d\textbf{r}_0.
    \label{champ_intermediaire}
\end{equation}
The propagation of the reflected wave-field from the sample plane $\Sigma_{\rho}$ to the \al{detector plane $\Sigma_d$} can be modelled by the output impulse response  $h_\textrm{out}(\textbf{r}_d,\textbf{r})$ between a point in the sample plane at coordinate $\textbf{r}$ and a \al{pixel of the camera in the detector plane $\Sigma_d$} at coordinate $\textbf{r}_{\rev{d}}$ (Fig. $\ref{fig1}$). This impulse response can be used to express the electric field $ E_{\al{d}}(\textbf{r}_{\al{d}},\alex{\omega})$ in the detector plane coming from the sample arm: 
\begin{equation}
    E_{\al{d}}(\textbf{r}_{\al{d}}\alex{,\omega}) = \int_{\Sigma_{\rho}} h_\textrm{out}(\textbf{r}_{\al{d}},\textbf{r}) E_{r}(\textbf{r}\alex{,\omega}) d\textbf{r},
    \label{champ_echantillon}
\end{equation}
where the integral is performed over the sample plane $\Sigma_{\rho}$.
By reinjecting Eq.~\ref{champ_intermediaire} into Eq.~\ref{champ_echantillon}, we obtain~:
\begin{equation}
    E_{\al{d}}(\textbf{r}_{\rev{d}}\alex{,\omega}) =  \int_{\Sigma_{0}  }\int_{  \Sigma_{\rho}}  h_\textrm{out}(\textbf{r}_{\al{d}},\textbf{r}) \rho(\textbf{r})  h_\textrm{in}(\textbf{r},\textbf{r}_0) E_{0}(\textbf{r}_0\alex{,\omega}) d\mathbf{r} d \mathbf{r}_0.
    \label{champ_echantillon2}
\end{equation}
In the following, for sake of simplicity, we will also assume a magnification of -1 between the source and the conjugated focal planes.
Under an isoplanatic hypothesis, this means that the impulse response satisfies $h_\textrm{in/out}(\mathbf{r},\mathbf{r}')=h\rev{_\textrm{in/out}}(\textbf{r}+\textbf{r}')$.
The expression of the output sample electric field can then be simplified as follows:
\begin{equation}
    E_{\al{d}}(\textbf{r}_{\al{d}}\alex{,\omega}) =  \int_{\Sigma_{0}  }\int_{  \Sigma_{\rho}}  h_{\rev{\textrm{out}}}(\textbf{r}_{\al{d}}+\textbf{r}) \rho(\textbf{r})  h_{\rev{in}}(\textbf{r}+\textbf{r}_0) E_{0}(\textbf{r}_0\alex{,\omega}) d\mathbf{r}_{\rev{d}} \rev{d}\mathbf{r}_0.
    \label{champ_echantillon3}
\end{equation}

\subsection{Standard incoherent image}

\alex{The standard incoherent image $ I_{s}(\textbf{r}_{\rev{d}})$ provided by a conventional microscope corresponds to the time integration of the square modulus of the time-dependent electric wave-field $e_{\al{d}}(\textbf{r}_{\rev{d}}\alex{,t})=\int d\omega E_{\al{d}}(\textbf{r}_{\al{d}},\omega)e^{j\omega t}$:
\begin{equation}
   I_{s}(\textbf{r}_{\al{d}})= \lim_{T\rightarrow \infty} \frac{1}{T} \int_0^T dt |e_{\al{d}}(\textbf{r}_{\al{d}},t)|^2 .
\end{equation}
Using Parseval's identity, this last expression can be rewritten by means of an integral over frequency:
\begin{equation}
   I_{s}(\textbf{r}_{\al{d}})= \frac{1}{\Delta \omega} \int_{\omega_-}^{\omega_+} d\omega |E_{\al{d}}(\textbf{r}_{\al{d}},\omega) |^2,
\end{equation}
with ${\omega_\pm}=\omega_c \pm \Delta \omega/2$ and $\omega_c$ the central frequency of the light source. For a sufficiently large bandwidth, the previous integral can be replaced by an ensemble average such that:
\begin{equation}
   I_{s}(\textbf{r}_{\al{d}})= \left \langle  |E_{\rev{d}}(\textbf{r}_{\al{d}},\omega) |^2\right \rangle,
\end{equation}
where the symbol $\langle \cdots \rangle$ accounts for the ensemble average over different realizations of the incident wave-field $E_{0}$.}

\alex{For incoherent imaging, the \rev{light source} is assumed to be both temporally and spatially incoherent\alex{, such that 
\begin{equation}
   \langle E_{0}(\textbf{r}_0,\omega) E^{*}_{0}(\textbf{r'}_0,\omega\rev{'}) \rangle = I_{0} (\textbf{r}_0) \delta (\textbf{r}_0-\textbf{r'}_0)\delta (\omega-\omega'),
    \label{incoherence_spatiale}
\end{equation}
where }$\delta $ is the Dirac distribution and $I_{0}(\textbf{r}_0)$ is the source intensity distribution.} \alex{The spatial incoherence of the \rev{source} yields to the following expression for \alex{$ I_{s}$}~:
\begin{align}
    I_{s}(\textbf{r}_{\al{d}} ) =& \int_{\Sigma_{\rho}} \int_{\Sigma_{\rho}}  d\textbf{r'} d\mathbf{r} h_{\rev{\textrm{out}}}(\textbf{r}_{\al{d}}+\textbf{r}) h_{\rev{\textrm{out}}}^*(\textbf{r}_{\al{d}}+\textbf{r'}) \rho(\textbf{r})  \rho^*(\textbf{r'}) W(\mathbf{r},\mathbf{r'}).
    \label{image_conventionnelle0}
\end{align}
\rev{$W(\mathbf{r},\mathbf{r'})$ is the cross-spectral density of the incident wave-field: 
\begin{equation}
W(\mathbf{r},\mathbf{r'})=  \langle E_{i}(\textbf{r},\omega) E^{*}_{i}(\textbf{r'},\omega) \rangle=\int_{\Sigma_0}  d\textbf{r}_0  h_{\rev{\textrm{in}}}(\textbf{r}_0+\mathbf{r})h_{\rev{\textrm{in}}}^*(\textbf{r}_0+\mathbf{r'})  I_{0} (\textbf{r}_0).
\end{equation}
To go further, one can first assume an homogeneous source distribution: $I_{0} (\textbf{r}_0) \equiv I_{0}$. Second, the aberrations of the imaging system are supposed to only induce phase distortions on the incident and reflected wave-fronts. Third, the input numerical aperture is assumed to match with the numerical aperture of the microscope objective. Under these assumptions, the incident light $E_i$ can then be assumed as spatially incoherent: $ W(\mathbf{r},\mathbf{r'}) \simeq I_0 \delta(\mathbf{r}-\mathbf{r'}) $.} \rev{Similarly to a conventional microscope in transmission~\cite{Hopkins1950,Sheppard1977,Goodman,Born}, the reflected} incoherent intensity $I_s$ can then be finally expressed as follows,} 
\begin{align}
    I_{s}(\textbf{r}_{\al{d}}) &=I_{0} \int |h_{\rev{\textrm{out}}}(\textbf{r}_{\al{d}} + \textbf{r})|^{2} |\rho(\textbf{r})|^{2}  d\textbf{r} 
    \nonumber \\
    &=I_{0} \left [ |h_{\rev{\textrm{out}}}|^{2} *  |\rev{\bar{\rho}}|^{2} \right ] (\textbf{r}_{\al{d}} ).
    \label{image_conventionnelle}
\end{align}
The symbol $*$ stands for the convolution product and $\rev{\bar{\rho}}(\textbf{r})=\rho(-\textbf{r})$. \rev{The incoherent image can thus be expressed as the convolution between the squared sample reflectivity, $|\rev{\bar{\rho}}|^2$, and the incoherent PSF, $|h_{\rev{\textrm{out}}}|^2$. Not surprisingly, we retrieve the fundamental property that a conventional microscope under incoherent illumination is equivalent to a scanning microscope\rev{~\cite{Zeitler1970,Welford1972,Sheppard1986}}.}  

To see the impact of aberrations, one can analyze the image in the Fourier domain: $\mathcal{I}_s(\mathbf{u})=\int d\mathbf{r}_s I_s(\mathbf{r}_s) \exp\left ( i 2 \pi \mathbf{u}.\mathbf{r}_s \sin \alpha /\lambda f   \right) $, with $\mathbf{u}$ the dual coordinate in the pupil plane. In the Fourier domain, Eq.~\ref{image_conventionnelle} can be written as follows~:
\begin{equation}
   \mathcal{I}_s(\mathbf{u}) = I_{0} \left[ \left( \mathcal{H}_\textrm{out} \otimes \mathcal{H}_\textrm{out} \right) \times \left( \mathcal{P} \otimes \mathcal{P} \right) \right] (\textbf{u}),
    \label{image_conventionnelle_Fourier}
\end{equation}
where $\otimes$ stands for the cross-correlation product. \rev{$\mathcal{H}_\textrm{out}$} and $\mathcal{P}$ are the inverse Fourier transforms of \rev{$h_\textrm{out}$} and \rev{$\bar{\rho}$}. The auto-correlation of the amplitude transfer function, \rev{$ \mathcal{H}_\textrm{out} \otimes \mathcal{H}_\textrm{out}$}, is the well-known optical transfer function (OTF)\rev{~\cite{Hopkins1950,Goodman,Born}}. As we will see further, aberrations behave like a spatial frequency filter given by the OTF for incoherent imaging. The loss of the high spatial frequencies of the object is illustrated by Fig.~\ref{fig2}\al{(g)} that shows the computed incoherent image in the conditions of the experiment depicted by Fig.~\ref{fig1}. The agreement with the experimental result (Fig.~\ref{fig2}\al{(c)}) is excellent. 

\subsection{\rev{FFOCT image}}

The FFOCT image is obtained by extracting the interference term \alex{$F(\mathbf{r}_{\al{d}})=\langle E_{\al{d}}(\textbf{r}_{\al{d}},\omega) E^{*}_{ref}(\textbf{r}_{\al{d}},\omega) \rangle$} between the reflected wave-fields coming from the sample and reference arms \alex{in Fig.~\ref{fig1}, under a fully incoherent illumination (Eq.~\ref{incoherence_spatiale}).} Light propagation in the reference arm is described by the impulse response $h_{\rev{\textrm{ref}}}$. In the ideal case, $h_{\rev{\textrm{ref}}}$ is an Airy disk whose radius $\delta_{\rev{\textrm{ref}}}$ is inversely proportional to the \rev{input numerical aperture}:\all{
\begin{equation}
\label{delta}
\delta_{\rev{\textrm{ref}}} \simeq 1.22 \lambda /(2 \alex{\sin \alpha_{\rev{\textrm{in}}}}).
\end{equation}}
In the Fourier domain, the associated pupil function $\mathcal{H}_{\rev{\textrm{ref}}}$ is constant and equal to 1 over the \rev{input} pupil aperture: 
\begin{equation}
    \mathcal{H}_{\rev{\textrm{ref}}}(\textbf{u})=\rev{\mathbf{1}}_{|\textbf{u}|< \rev{u_\textrm{in}} },
    \label{pupil_function_ref}
\end{equation}
\rev{with $u_\textrm{in}={{\sin \alpha_{\textrm{in}}}/{\sin \alpha}}$.} Considering the reference mirror as perfect ($\rho(\mathbf{r})\equiv 1$), the reference wave-field can be obtained by replacing $h_\textrm{in/out}$ with $h_\textrm{ref}$ in Eq.~\ref{champ_echantillon3}: 
\begin{eqnarray}
    E_\textrm{ref}(\textbf{r}_{\al{d}}\alex{,\omega})& = & \int_{\Sigma_{0}  }\int_{  \Sigma_\textrm{ref}}  h_{\rev{\textrm{ref}}}(\textbf{r}_{\al{d}}+\textbf{r}')  h_{\rev{\textrm{ref}}}(\textbf{r}'+\textbf{r'}_0) E_{0}(\textbf{r'}_0\alex{,\omega}) d\mathbf{r}' d \mathbf{r'}_0\\
    & =& \int_{\Sigma_{0}  } h_{\rev{\textrm{ref}}}(\textbf{r}_{\al{d}}-\textbf{r'}_0) E_{0}(\textbf{r'}_0\alex{,\omega}) d\textbf{r'}_0 
    \label{champ_reference}
\end{eqnarray}
The reference wave-field is a diffraction-limited replica of the source wave-field.

A theoretical expression for the FFOCT image, $F(\mathbf{r}_{\al{d}})$, can be expressed by combining Eqs.~\ref{champ_echantillon3} and \ref{champ_reference}: 
\begin{eqnarray}
 F(\mathbf{r}_{\al{d}})  &= &\iiint h_{\rev{\textrm{out}}}(\textbf{r}_{\al{d}}+\textbf{r}) \rho(\textbf{r}) h_{\rev{\textrm{in}}}(\textbf{r}+\textbf{r}_0) h^{*}_{\rev{\textrm{ref}}}(\textbf{r}_{\al{d}}-\textbf{r}'_0) \nonumber  \\
 &\times & \langle E_{0}(\textbf{r}_0,\omega) E^{*}_\textrm{0}(\textbf{r}'_0,\omega) \rangle
 d\textbf{r} d\textbf{r}_0 d\textbf{r}'_0.
    \label{terme_interference}
\end{eqnarray}
Under a fully incoherent illumination (Eq.~\ref{incoherence_spatiale}), this last expression can be simplified as follows: 
\begin{align}
     F(\mathbf{r}_{\al{d}}) 
    =I_{0} \left[ h_{\rev{\textrm{out}}} \times \left ( h_{\rev{\textrm{in}}} * h^{*}_{\rev{\textrm{ref}}} \right ) \right] * \rev{\bar{\rho}} (\textbf{r}_{\al{d}}).
    \label{terme_interference2}
\end{align}
In the Fourier domain, the previous equation becomes: 
\begin{equation}
   \mathcal{F}(\mathbf{u}) = I_{0} \left \lbrace \left[ \mathcal{H}_{\rev{\textrm{out}}} * \left( \mathcal{H}_{\rev{\textrm{in}}} \times \al{\rev{\bar{\mathcal{H}}}^{*}_{\rev{\textrm{ref}}}} \right) \right] \times \mathcal{P} \right \rbrace (\textbf{u}).
    \label{terme_interference_Fourier}
\end{equation}
\al{with $\rev{\bar{\mathcal{H}}}_{\rev{\textrm{ref}}}(\mathbf{u})=\rev{\mathcal{H}}_{\rev{\textrm{ref}}}(-\mathbf{u})$.} In the Linnik configuration, the microscope objectives are identical in the two arms. The \rev{input} numerical aperture is thus the same in both arms, hence the support of $\mathcal{H}_{\rev{\textrm{in}}}$ is contained in the support of $\mathcal{H}_\textrm{ref}$ (Eq.~\ref{pupil_function_ref}). Therefore, $\mathcal{H}_{\rev{\textrm{in}}} \times {\rev{\bar{\mathcal{H}}}^{*}_{\rev{\textrm{ref}}}} = \mathcal{H}_{\rev{\textrm{in}}}$. Equation~\ref{terme_interference_Fourier} can be rewritten as follows: 
\begin{equation}
   \mathcal{F}(\mathbf{u}) = I_{0} \left \lbrace \alex{\mathcal{H}_2} \times \mathcal{P} \right \rbrace (\textbf{u}),
    \label{terme_interference_Fourier2}
\end{equation}
where $ \mathcal{H}_2 (\mathbf{u})$ is the confocal ATF:
\begin{equation}
\label{H2}
    \mathcal{H}_2 (\mathbf{u})=\left[  \mathcal{H}_{\rev{\textrm{out}}} * \mathcal{H}_{\rev{\textrm{in}}}  \right] (\textbf{u}).
\end{equation}
\rev{Note that expressions similar to Eq.~\ref{terme_interference_Fourier2} have been derived for three-dimensional scattering media in the monochromatic regime~\cite{Marks2009} and, more recently, for the time-gated image~\cite{Tricoli2019}. Equation~\ref{terme_interference_Fourier2}} thus holds in a more general context than a planar object. Going back to the spatial domain, one finally obtains~: 
\begin{align}
   F(\mathbf{r}_{\al{d}})
    &=I_{0}  \left [\rev{h_2}  * \rev{\bar{\rho}} \right ](\textbf{r}_{\al{d}}),
    \label{terme_interference3}
\end{align}
\al{where \rev{$h_2=h_{\rev{\textrm{out}}} \times h_{\rev{\textrm{in}}}$} is the coherent confocal PSF~\cite{Sentenac2018}. \rev{While the output impulse response $h_{\rev{\textrm{out}}}$ grasps the effect of the numerical aperture of the microscope objective and the aberrations undergone by the reflected wave-front, the input impulse response $h_{\rev{\textrm{in}}}$ accounts for the aberrations and the partial coherence exhibited by the incident wave-field. Equation~\ref{terme_interference3} thus represents a general model for partially coherent time-gated FFOCT, ranging from the coherent case of an incident plane wave ($\mathcal{H}_\textrm{in}(\mathbf{u})=\delta(\mathbf{u})$, with $\delta$ the Dirac distribution) to a spatially incoherent illumination scheme ($\mathcal{H}_\textrm{in}\equiv\mathcal{H}_\textrm{out}$).}} 

\rev{Equation~\ref{terme_interference3} is remarkable in several ways. First, as already pointed out in the frequency domain~\cite{Sheppard1977,Marks2009} and by virtue of the principle of reciprocity~\cite{Sheppard1986,Sentenac2018}, FFOCT is equivalent to a time-gated confocal image}, that can be obtained in optical coherence microscopy if a pinhole was placed in front of the detector~\cite{Kempe1996} or in fiber-based OCT systems where single-mode optical fiber serves as a pinhole aperture for both illumination and collection of light from the sample~\cite{Izatt2008}. Indeed, for a point-like souce in Fig.~\ref{fig1}, $E_0(\mathbf{r}_0)=I_0\delta(\mathbf{r}_0-\mathbf{r}_{\rev{d}})$, the time-gated confocal wave-field $E_{\al{d}}(\mathbf{r}_{\al{d}})$ (Eq.~\ref{champ_echantillon3}) exactly \rev{matches the FFOCT signal obtained under a spatially incoherent illumination (Eq.~\ref{terme_interference3}). In the latter case, this is the spatial incoherence of the light source that acts as a physical confocal pinhole~\cite{Karamata2004}.}

Second, the comparison of Eq.~\ref{terme_interference3} with the \rev{incoherent image of the conventional microscope} (Eq.~\ref{image_conventionnelle}) shows the benefit of low coherence interferometry. While incoherent imaging yields an image of the square norm $|\rho|^2$ of the reflectivity with an \al{incoherent imaging PSF \rev{$|h_\textrm{out}|^2$}}, FFOCT provides a coherent image of the sample reflectivity $\rho$ with the amplitude PSF \rev{$h_2$}. \rev{Compared to the incoherent image, \rev{Eq.~\ref{terme_interference_Fourier2}} shows that \alex{phase} aberrations \alex{($|\mathcal{H}_{\rev{\textrm{out}}}(\mathbf{u})|\equiv 1$)} do not filter the high spatial frequencies of the object. Its frequency components are nevertheless dephased between each other because of phase aberrations. Figures~\ref{fig2}(j) and (f) illustrate this fact by showing the FFOCT image of the object that would be obtained for a coherent and spatially incoherent illumination, respectively, under the conditions of the experiment depicted in \al{Fig.~\ref{fig2}}. The comparison with the incoherent image [Fig.~\ref{fig2}\al{(g)}] demonstrates the robustness of FFOCT with respect to low-order aberrations such as defocus. This striking result will be discussed in details in Sec.~5. The comparison between Figs.~\ref{fig2}(j) and (f) shows the slight gain of resolution provided by a spatially incoherent illumination. The confocal PSF $h^2_{\rev{\textrm{out}}}$ is actually thinner than the coherent PSF $h_{\rev{\textrm{out}}}$. The main reason for the robustness of FFOCT to defocus is nevertheless its coherent feature. \rev{Besides resolution, the virtual confocal pinhole drastically improves the image contrast by spatially filtering a large part of multiply-scattered photons~\cite{Karamata2004,Marks2009} taking place ahead of the focal plane and arriving in the same time gate as singly-scattered photons.}}
 
Before a more quantitative study about the effect of aberrations in FFOCT, it is important to first derive its diffraction-limited resolution and compare it with the incoherent image provided by a standard microscope. 

\section{Diffraction-limited resolution}

\rev{Under a spatially incoherent illumination and in absence of aberrations, the input and output pupil functions coincide with the microscope pupil function $\mathcal{H}_0(\mathbf{u})$, such that
\begin{equation}
\label{H0}
     \mathcal{H}_{\rev{\textrm{in/out}}} \equiv  \mathcal{H}_0=\mathbf{1}_{|u|<1}.
\end{equation}
In this ideal case,} the incoherent OTF and the FFOCT ATF are strictly equal
\begin{equation}
\label{equal}
\left [ \mathcal{H}_0 * \mathcal{H}_0 \right] (\mathbf{u})  \equiv \left [ \mathcal{H}_0 \otimes \mathcal{H}_0 \right] (\mathbf{u}) =\Lambda(\mathbf{u}) ,
\end{equation}
where
\begin{equation}
\Lambda(\mathbf{u})=
\frac{2}{\pi} \left [ \arccos (u/2) - (u/2)\sqrt{1-\left(u/2 \right )^2} \right ]
   \label{OTF0}
\end{equation} 
for \alex{$|u|<2$} and zero elsewhere~\cite{Goodman}.
\begin{figure*}[htbp]
\centering
\includegraphics[width=\linewidth]{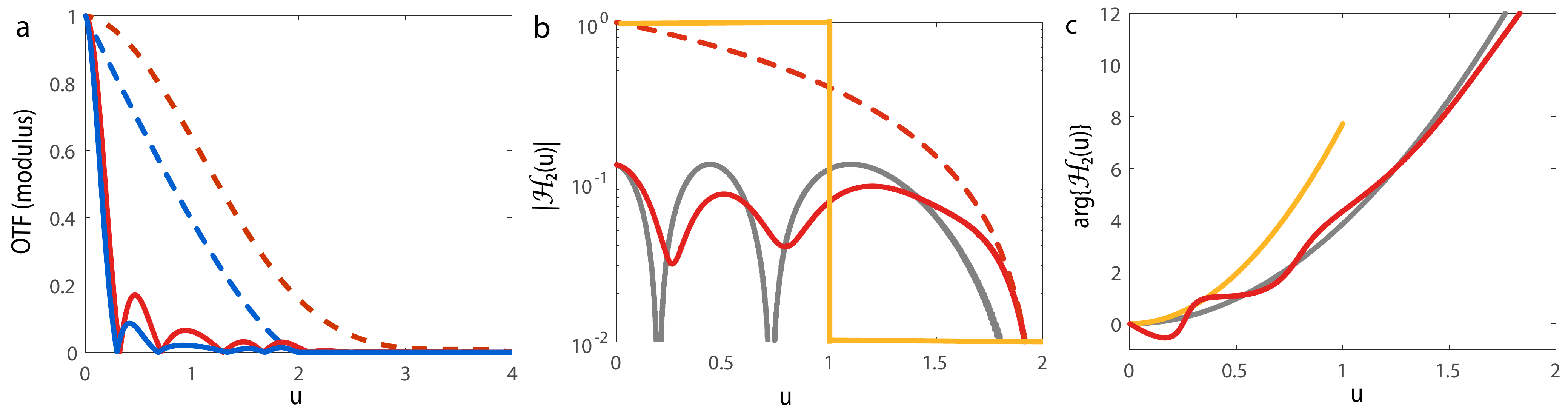}
\caption{Optical and amplitude transfer functions associated with the experiment depicted in Fig.~\ref{fig1}. (a) The \al{modulus of the OTF (referred to as modulation transfer function in the literature)} is shown for: (\textit{i}) FFOCT in presence (\alex{$[\mathcal{H}_2 \otimes \mathcal{H}_2] (\mathbf{u}) $}, red continuous line) and absence \alex{($[\Lambda  \otimes \Lambda](\mathbf{u}) $, red dashed line)} of defocus, and (\textit{ii}) for incoherent imaging in presence (\alex{$[\mathcal{H}  \otimes \mathcal{H}] (\mathbf{u}) $}, blue continuous line) or absence \alex{($\Lambda(\mathbf{u}$) , blue dashed line)} of defocus. (b) The modulus of the exact confocal ATF (\alex{$|\mathcal{H}_2 (\mathbf{u})|$}, red line) is compared to its analytical approximation (\alex{grey line}, Eq.~\ref{approx}), to its value in absence \alex{of} aberrations (\alex{$\Lambda(\mathbf{u})$, red dashed line}) and \alex{to the standard coherent ATF (\alex{$|\mathcal{H}_0 (\mathbf{u})|$}, yellow line)}. Note that the y-axis is in log-scale. (c) The phase of the exact confocal ATF (\alex{$\mathcal{H}_2(\mathbf{u})$}, red line) is compared to its analytical approximation (\rev{gray} line, left term of Eq.~\ref{approx}) and to the original defocus phase shift exhibited by the pupil function (yellow line, Eq.~\ref{defocus_pupil}).}
\label{fig4}
\end{figure*}
$\Lambda(\mathbf{u})$ is plotted in Fig.~\ref{fig4}(a). Both the OTF and ATF are shown to extend to a frequency that is twice the coherent cutoff frequency $u_c=1$ of $\mathcal{H}_0$. This should not be taken to imply that both a conventional microscope and a FFOCT apparatus have the same resolving power~\cite{Goodman}. A major flaw lies in the direct comparison of the cutoff
frequencies in the two cases. Actually, the two are not directly comparable, since the ATF cutoff determines the maximum spatial frequency component
of the FFOCT field amplitude while the OTF cutoff determines
the maximum spatial frequency component of incoherent image intensity. Surely any direct comparison of the two systems must be in terms of the same observable quantity, \textit{i.e} the image
intensity. To do so, we will study, in the following, the intensity of the FFOCT signal, $I_f(\mathbf{r}_{\al{d}})=|F(\mathbf{r}_{\al{d}})|^2$. In the Fourier domain, the spatial frequency spectrum of the FFOCT image is thus given by:
\begin{equation}
   \mathcal{I}_F(\mathbf{u}) = \left \lbrace \alex{\mathcal{H}_2} \times \mathcal{P} \right \rbrace \otimes \left \lbrace \alex{\mathcal{H}_2}  \times \mathcal{P} \right \rbrace (\textbf{u}).
    \label{terme_interference_Fourier3}
\end{equation}
While the finite pupil aperture behaves as a spatial frequency filter, \rev{$\mathcal{H}_\textrm{out}\otimes \mathcal{H}_\textrm{out} $}, in a conventional microscope (Eq.~\ref{image_conventionnelle_Fourier}), its effect depends on the nature of the object in FFOCT (Eq.~\ref{terme_interference_Fourier3}).

Nevertheless, the diffraction-limited resolution can be derived by considering the case of a point-like bead, as in the experiment depicted in Fig.~\ref{fig3}. In that case, $\mathcal{P}(\mathbf{u})\equiv 1$. The corresponding incoherent and FFOCT spectra, $\mathcal{I}^{(o)}_S(\mathbf{u})$ and $\mathcal{I}^{(o)}_F(\mathbf{u})$, then simplify into
\begin{equation}
\mathcal{I}^{(o)}_S(\mathbf{u})  =  \left[  \mathcal{H}_{\rev{\textrm{out}}} \otimes  \mathcal{H}_{\rev{\textrm{out}}} \right ]   (\textbf{u})
\end{equation}
and 
\begin{equation}
\mathcal{I}^{(o)}_F(\mathbf{u}) = \left[  \mathcal{H}_{\rev{\textrm{out}}} *  \mathcal{H}_{\rev{\textrm{in}}} \right ] \otimes \left[  \mathcal{H}_{\rev{\textrm{out}}} *  \mathcal{H}_{\rev{\textrm{in}}} \right ]  (\textbf{u}).
    \label{terme_interference_Fourier4}
\end{equation}
In absence of aberrations [Eq.~\ref{H0}], the corresponding FFOCT spectrum  $\mathcal{I}^{(o)}_F(\mathbf{u})$ (Eq.~\ref{terme_interference_Fourier3}) simply becomes $ [ \Lambda \otimes \Lambda ] (\textbf{u})$ while its incoherent counterpart scales as $\Lambda(\mathbf{u})$. The radial dependence of $\mathcal{I}^{(o)}_S(\mathbf{u})$ and $\mathcal{I}^{(o)}_F(\mathbf{u})$ in these ideal conditions are displayed in Fig.~\ref{fig4}(a). While the FFOCT image spans over a spatial frequency range $\Delta u_F = 4$, the incoherent intensity spectrum displays a typical width $\Delta u_S = 2$. \all{The theoretically achievable transverse resolution in FFOCT, 
\begin{equation}
\label{confoc_res}
\delta^{(o)}_F \sim 1.22 \lambda/(4 \sin \alpha),
\end{equation}
is thus reduced by a factor 2
compared to the standard diffraction-limited resolution \rev{$\delta_0=1.22 \lambda/(2\sin \alpha)$}~\cite{Sentenac2018}.} This explains, in part, the much thinner PSF exhibited by the FFOCT image compared to the incoherent one for a point-like bead (see Fig.~\ref{fig3}). 
The corresponding intensity PSF can be deduced from the inverse Fourier transform of Eq.~\ref{terme_interference_Fourier4}. The experimental PSF at zero defocus [Fig.~\ref{fig3}(d1)] is in good agreement with our theoretical prediction [Fig.~\ref{fig3}(e1)]. Note that the gain in resolution compared to the incoherent PSF [Fig.~\ref{fig3}(c1)] is experimentally larger than 2 because the incoherent PSF also suffers from scattering events taking place ahead of the focal plane. These multiply-scattered photons\rev{, neglected by our model,} widen the incoherent PSF whereas the coherent time gating operation allows us to get rid of them in FFOCT. 

\section{Robustness to defocus}

Now that the diffraction-limited resolution has been rigorously derived, the manifestation of aberrations in FFOCT can be investigated and compared to the other imaging modes. To fully explain the results shown in Figs.~\ref{fig2} and \ref{fig3}, we will first consider the case of a simple defocus. Because the impact of aberrations is object-dependent in FFOCT, the case of a point-like target (Fig.~\ref{fig3}) will be first considered before tackling a coherent object such as a resolution target. (Fig.~\ref{fig2}). 

\subsection{Defocus \rev{in a conventional microscope}}

Mathematically, the output pupil function can be expressed as follows for an error of focus~\cite{Gu1992,Sheppard1991}:
\begin{equation}
   \mathcal{H}_{\rev{\textrm{out}}}(\mathbf{u},z) = \exp \left (i W |\mathbf{u}|^2/2 \right )    \mathcal{H}_0(\mathbf{u}),
    \label{defocus_pupil}
\end{equation}
where the path length error,
\begin{equation}
    W= \frac{8 \pi  }{\all{n}\lambda} z  \sin^2 (\alpha/2),
\end{equation} 
is a convenient indicator of the severity of the focusing error, \all{$n$, the optical index in the surrounding medium} and $z$, the defocus distance from the focal plane. \alex{The modulus and phase of $ \mathcal{H}_{\rev{\textrm{out}}}(\mathbf{u},z)$ corresponding to the experiment depicted in Fig.~\ref{fig1} are plotted as a reference in Fig.~\ref{fig4}(b) and (c), respectively.}

An analytical expression of the OTF has been derived in a recent work~\cite{Liang2017}. In the geometric limit ($W>>1$), it simplifies into:
\begin{equation}
\label{geometric}
\mathcal{I}^{(o)}_S(\mathbf{u},z)=\left [ \mathcal{H}_{\rev{\textrm{out}}} \stackrel{\mathbf{u}}{\otimes} \mathcal{H}_{\rev{\textrm{out}}}\right ](\mathbf{u},z) \simeq \frac{2J_1 (W|\mathbf{u}|) }{W|\mathbf{u}|} .
\end{equation}
We can verify that this is precisely the OTF predicted by geometrical optics, that is to say the geometrical projection of the exit pupil into the image plane, and therefore the point-spread function $|h_{\rev{\textrm{out}}}|^2$ should be uniformly bright over a disk of radius $\delta_{\infty}$ and zero elsewhere,
\begin{equation}
\label{extension}
   \left |h_{\rev{\textrm{out}}}(\mathbf{r},z) \right |^2=\mathbf{\rev{1}}_{|\textbf{r}|\leq \delta_\infty} ,
\end{equation}
with 
\begin{equation}
\label{dgeom}
\all{\delta_\infty \sim   \frac{z \sin \alpha}{\sqrt{n^2- \sin^2 \alpha }} .}
\end{equation}
The perimeter of this disk is superimposed to the incoherent images of the beads under defocus in Fig.~\ref{fig3}(c). As expected, the agreement between experiment and theory becomes better when we approach the geometrical limit. Again, the discrepancy between theory and experiment at small defocus is mainly due to the forward multiple scattering events \alex{induced by the agarose gel} outside of the focal plane. 

As illustrated by Eq.~\ref{geometric}, strong aberrations can cause the OTF to have negative values in certain bands of frequencies [see the secondary lobes in Fig.~\ref{fig4}(a)]. When the OTF is negative, image components at that spatial frequency undergo a contrast reversal~\cite{Goodman}; i.e., intensity peaks become intensity nulls, and vice versa. This effect is nicely retrieved on the incoherent image of the Siemens target both experimentally [Fig.~\ref{fig2}(\al{c})] and theoretically [Fig.~\ref{fig2}(\al{g})]. In both figures, the main contrast reversal is highlighted by a white dashed line. The local spatial frequency of the Siemens increases slowly when we approach its center. The local contrast of fringes is thus an indication of the value of the OTF at the corresponding spatial frequency. When the system is out of focus, a gradual attenuation of contrast and a number of contrast reversals are obtained for increasing spatial frequency, as predicted by Eq.~\ref{geometric}. 

\subsection{Defocus in FFOCT}

\subsubsection{\rev{Amplitude transfer function for a spatially-incoherent illumination}}
With regards to FFOCT, one can benefit from previous works that have investigated the 3D PSF in coherent confocal microscopy~\cite{Sheppard1991, Gu1992}. \rev{Under spatially incoherent illumination,} the corresponding ATF, $\mathcal{H}_2(\mathbf{u},z)=\left [\mathcal{H}_{\rev{\textrm{out}}}*\mathcal{H}_{\rev{\textrm{out}}}\right ](\mathbf{u},z)$, can actually be decomposed by means of a Fourier decomposition along $z$, such that
\begin{equation}
    \alex{\mathcal{H}_2(\mathbf{u},z)=\int ds \hat{\mathcal{H}}_2(\mathbf{u},s) \exp \left (i 2 \pi s z \right )}.
\end{equation}
An analytical solution can be derived for each Fourier component of $H_2(\mathbf{u},z)$:
\begin{equation}
 \alex{\hat{\mathcal{ H}}}_2(\mathbf{u},s)=\left \{
\begin{array}{ll}
1 & \, \mbox{for } \frac{u^2}{4} \leq s \leq 1-u\left ( 1-\frac{u}{2} \right )  \\
\frac{2}{\pi} \arcsin \left (\frac{1-s}{u\sqrt{s-u^2/4}}\right) & \,  \mbox{for }1-u\left (1-\frac{u}{2} \right ) \leq s  \leq 1  \\
0 & \, \mbox{otherwise.}
\end{array} 
\right. 
\end{equation}
The ATF $\mathcal{H}_2(\mathbf{u},z)$ does not have any analytical expression but it can be approached by restricting its spatial frequency spectrum from $u^2/4$ to $1-u\left ( 1-u/2 \right ) $. It yields the following expression:
\begin{equation}
    \mathcal{H}_2(\mathbf{u},\alex{z}) \simeq \frac{\exp \left( i W u^2 / 4 \right )}{W} \left \lbrace \exp \left [i W (1-u/2)^2 \right ] -1 \right \rbrace \rev{\mathbf{1}}_{|\textbf{u}|<\alex{2}} 
    \label{approx}
\end{equation}
This approximate analytical expression of $\mathcal{H}_2$ is compared to its exact value in Fig.~\ref{fig4}(b) and (c), using the parameters of the experiment depicted in Fig.~\ref{fig1}. Albeit only approximate, Equation~\ref{approx} enables a physical interpretation of the confocal ATF. First the left term exhibits a quadratic phase shift that is two times smaller than the pupil function $\mathcal{H}_{\rev{\textrm{out}}}$. This phase shift is actually a good approximation of the phase of $\mathcal{H}_2$ [see Fig.~\ref{fig4}(c)]. Surprisingly, the confocal filter thus seems to reduce the defocusing effect. 

\subsubsection{\rev{Point-spread function for a spatially-incoherent illumination}}

Figure~\ref{fig3}(e) displays the theoretical PSFs computed in the conditions of the experiment described in Fig.~\ref{fig3}(a). \all{For a defocus smaller than the depth-of-field ($W<4\pi$, i.e $z<n\lambda /[2\sin^2 (\alpha/2)] \sim 4$ $\mu$m in Fig.~\ref{fig2}), the phase shift displayed by $\mathcal{H}_2$ remains inferior to $\pi$ and its extended angular aperture ($\Delta u=2$) gives rise to a gain in resolution compared to standard imaging [see Fig.~\ref{fig3}(e1) and (e2)].} 
\all{For a defocus larger than the depth-of-field ($W>4\pi$, i.e $z>n\lambda /[2\sin^2 (\alpha/2)]$), the reduction of the defocusing effect shown by Eq.~\ref{approx} and the larger extension exhibited by $\mathcal{H}_2$ counteract each other.} The confocal coherent PSF, $h_2(\mathbf{r},z)=h_{\rev{\textrm{out}}}^2(\mathbf{r},z)$, tends towards the incoherent PSF in the geometric limit, namely a disk of radius $\delta_\infty$ (Eq.~\ref{extension}), and the corresponding spectrum, $ \mathcal{I}^{(o)}_F(\mathbf{u})$, towards the Airy disk (Eq.~\ref{geometric}). \all{This geometric regime is reached in Fig.~\ref{fig3}(e3) and (e4) beyond $z=4$ $\mu$m: The confocal PSF then displays the same geometric extension as in incoherent imaging (Eq.~\ref{dgeom}).} \rev{Note that the different behavior of FFOCT for a defocus within or beyond the Rayleigh range has already been noticed by Marks \textit{et al.}~\cite{Marks2009} assuming a Gaussian distribution for the pupil functions. It thus seems a general feature of FFOCT.}

The Fresnel rings displayed by \alex{theoretical} FFOCT PSFs in Fig.~\ref{fig3}(e) are a direct consequence of the amplitude modulation of $\mathcal{H}_2$ [Figs.~\ref{fig2}(h) and \ref{fig4}(b)], reminiscent of a Fresnel zone plate. The right term in Eq.~\ref{approx} accounts for this modulation. The ATF $\mathcal{H}_2$ is made of $N=E [ W/(2\pi)]$ Fresnel zones (where $E(x)$ is the integer part function). The contribution of each Fresnel zone to the PSF at focus is alternatively positive or negative. A constructive or destructive interference is obtained at the focus when $W/(2\pi)$ is close to be an odd or even integer, respectively.
 
This is confirmed by deriving the analytical expression of the confocal PSF on the optical axis ($\mathbf{r}=\mathbf{0}$)
\begin{equation}
\label{onaxisPSF}
h_2(\mathbf{0},z)=\exp \left (i u/2 \right ) \sin_c^2 \left( W/4 \right).
 \end{equation}
The evolution of the PSF amplitude $h_2(\mathbf{0},z)$ is plotted as a function of the defocus distance $z$ in Fig.~\ref{fig3}(b). As expected, the PSF cancels for defocus distances satisfying $W/4=m\pi$ with $m\in\mathbb{N}^*$. For instance, at $z=9$ $\mu$m, the ATF is made of $N\simeq 6$ Fresnel zone plates, hence the destructive interference displayed by the theoretical PSF in Fig.~\ref{fig3}(e4). On the contrary,  some particular values of defocus ($W/4=(m+1/2)\pi$) can lead to an odd number of Fresnel zones and a constructive interference right at $\mathbf{r}=\mathbf{0}$. The central lobe is then only limited by diffraction but this is, of course, at the cost of strong secondary lobes.

\all{The agreement between the experimental and theoretical PSFs in Fig.~\ref{fig3} is only qualitative.  First, at defocus smaller than the depth-of-field, the theoretical FFOCT PSF is thinner than the experimental one. Spherical aberrations due to the index mismatch between the Agar gel sample and the air microscope objective can account for it. At larger defocus, the spatial extension of the theoretical and experimental PSFs both scale as the geometrical prediction (Eq.~\ref{dgeom}). However, while the theoretical FFOCT PSF exhibits Fresnel rings characteristic of an out-of-focus [see Figs.~\ref{fig3}(e3) and (e4)], the experimental PSF displays a speckle-like feature [see Figs.~\ref{fig3}(d3) and (d4)]. Actually, our model reduces the coherence volume to a plane perpendicular to the optical axis, while, in reality, scattering events induced by the agarose gel can distort the incident and reflected wave-fronts, thereby leading a random-like deformation of the coherence plane. The study of the coherence plane deformation in scattering media is out-of-scope for this paper but it should be definitively addressed in the near future to accurately predict the imaging performance of FFOCT in biological tissues.}


\subsubsection{\rev{Point-spread function for a partially coherent illumination}}

\begin{figure*}
    \centering
    \includegraphics[width=\linewidth]{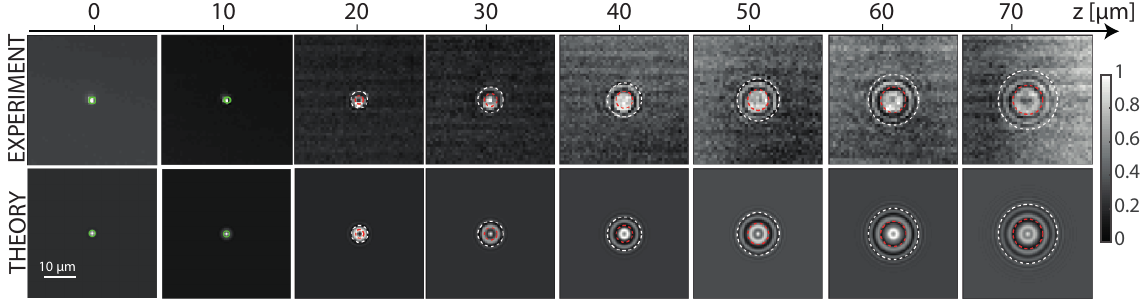}
    \caption{\all{FFOCT imaging of a gold \al{nano-bead} for various amounts of defocus. Experimental (top) and theoretical (bottom) images are displayed for different amount of defocus $z$. The spatial extension $\delta^{(o)}_F$  of the confocal PSF (green line, Eq.~\ref{confoc_res}) is superimposed on each image for $z < 20$ $\mu$m. The spatial extension $\delta_\infty$ of the input (red dashed line) and output (white dashed line) geometrical PSF (Eq.~\ref{dgeom}) is superimposed on each image for $z \geq 20$ $\mu$m. All images are normalized by their maximum.}}
    \label{figJM}
\end{figure*}


\all{To get a quantitative comparison with the developed model, an experiment similar to Ref.~\cite{Xiao_16_osa} has been performed by means of a commercial LLtech FFOCT system Light-CT Scanner~\cite{LLTech} ($\lambda \sim 626$ nm, $\Delta \lambda \sim 12$ nm). A single 80 nm-diameter gold nanobead placed on top of a coverslip is imaged through an immersion microscope objective ($ \sin \alpha=0.3$) over a large range of defocus ($z=0-70$ $\mu$m). The refractive index $n$ of the immersion oil is $n=1.515$. The experimental set up and procedure are described in Supplement 1. The experimental FFOCT images are displayed in the top panels of Fig.~\ref{figJM}. These images can be decomposed as the sum of the targeted nano-bead contribution and an homogeneous background due to a back-reflection induced by the coverslip on which is deposited the bead (see Supplement 1). The experimental images are thus fitted accordingly with the nano-bead image predicted by our FFOCT model plus a constant background of complex amplitude $B$. The K\"{o}hler illumination scheme used in the LLtech FFOCT system implies a partially coherent incident wave-field. \rev{This partial coherence can be taken into account by our model by considering an input numerical aperture smaller than the microscope objectives' one. Here the fit of experimental images yields $\sin \alpha_\textrm{in}=0.5\sin \alpha$.} The resulting theoretical images are displayed in the bottom panels of Fig.~\ref{figJM}. \rev{The experimental results and theoretical predictions are in good agreement.} The experimental PSF exhibits both the spatial extension and Fresnel rings predicted by our model. For a defocus smaller than the depth-of-field [here $ z< 20$ $\mu$m], the spatial extension of the PSF is shown to be close to the confocal theoretical resolution $\delta_F^{(0)}$ (Eq.~\ref{confoc_res}). For a defocus larger than the depth-of-field, the spatial extension of the FFOCT PSF is shown to tend towards the geometrical PSF (Eq.~\ref{dgeom}), here limited by the input numerical aperture. Note, however, that a slight disagreement subsists between experiment and theory, probably due to an imperfect modelling of the input illumination and also to residual aberrations induced by the imaging apparatus. } 

\all{Now that our theoretical model has been quantitatively validated by the latter experiment, the FFOCT imaging performance can now be investigated for an extended and coherent object.}

\subsubsection{\alex{Coherent object}}

As shown by Fig.~\ref{fig2}, FFOCT displays much better results than a \rev{conventional microscope} for a coherent object such as a resolution target. To explain the reason for this better performance, we have computed theoretically \al{three} images: the \rev{conventional} image [Eq.~\ref{image_conventionnelle_Fourier}, Fig.~\ref{fig2}\al{(g)}] \rev{and FFOCT images  (Eq.~\ref{terme_interference_Fourier3}) under spatially-coherent [Fig.~\ref{fig2}\al{(f)}] and incoherent [Fig.~\ref{fig2}(j)] illuminations}.  

A first remark is that the \al{theoretical confocal image [Fig.~\ref{fig2}\al{(j)}] and the standard incoherent image} [Fig.~\ref{fig2}\al{(g)}] are in \alex{good} agreement with the experimental results [Figs.~\ref{fig2}(b) and (c), respectively]. The contrast inversion, highlighted by a dashed white circle \alex{in Fig.~\ref{fig2}}, occurs at \alex{roughly} the same spatial frequency for experiment and theory. \alex{The reason for the slight discrepancy between the experimental and theoretical FFOCT images is the presence of residual aberrations in the experiment in addition to defocus. A possible origin is the deformation of the coherence plane that is not taken into account by our model.}

The second remark is that 
\al{the FFOCT images [Figs.~\ref{fig2}(\al{f}) and (j), respectively] show a much better resolution than the incoherent image [Fig.~\ref{fig2}(\al{g})].} This can be explained as follows: For incoherent imaging, the OTF is given by the autocorrelation of the pupil function and is thus independent on the spatial frequency spectrum of the object. On the contrary, for coherent imaging, the aberration pupil function is projected on the object's spectrum beforehand. The resolution target's spectrum is a decreasing function whereas the phase shift due to defocus increases with the spatial frequency $u$. As a consequence, a coherent image shall be less impacted by a defocus than the incoherent image whose OTF is uniformly impacted by aberrations over the whole numerical aperture. 

\al{The third remark is that} the FFOCT image is slightly better \rev{under a spatially-incoherent illumination} with a contrast inversion occuring at a \alex{slightly} larger spatial frequency cutoff \rev{than for an incident plane wave}. This gain \al{comes from} the reduction \al{by} a factor 2 of the parabolic phase shift \al{induced by} the virtual confocal pinhole in \rev{spatially-incoherent} FFOCT (Eq.~\ref{approx}). 

Now that our theoretical model has enabled us to account for all the seemingly contradictory results shown in this paper (Figs.~\ref{fig2}, \ref{fig3} \alex{and \ref{figJM}}) and in previous works~\alex{\cite{Xiao_16_osa}}, we now go beyond a simple defocus and extend our analysis to low-order aberrations in general.

\section{Low-order aberrations}

To investigate low-order aberrations, Zernike polynomials form a natural basis for analyzing wave-front aberrations in optical systems with circular pupils. As we will see now, the symmetry of the phase function has a strong impact on the manifestation of aberrations in FFOCT. As this symmetry is related to the the parity of the Zernike polynomial, we will investigate the manifestation of the aberration modes associated with an even and odd Zernike polynomial. \rev{In the following, for sake of simplicity, we will only consider the case of spatially-incoherent FFOCT ($\mathcal{H}_\textrm{in}=\mathcal{H}_\textrm{out}\equiv \mathcal{H}$). }

\subsection{\label{odd} Odd phase pupil function}

For an odd phase pupil function (odd Zernike polynomial, \textit{e.g} coma), $\mathcal{H}$ satisfies $\mathcal{H}(-\textbf{u})=\mathcal{H}^{*}(\textbf{u})$, hence $ [\mathcal{H} * \mathcal{H}] (\textbf{u}) \equiv [\mathcal{H} \otimes \mathcal{H}] (\textbf{u})$. In this case, the OTF of the incoherent image and the ATF of FFOCT are strictly equivalent:
\begin{equation}
    \mathcal{H}_2(\textbf{u})= [\mathcal{H} \otimes \mathcal{H}] (\textbf{u}).
\end{equation}
As for incoherent imaging, anti-symmetric aberrations behave like a spatial frequency filter in FFOCT. The support of the ATF spans from 0 to $u_c$, the effective spatial frequency cut-off. The latter quantity is here directly equal to the correlation width $u_{\mathcal{H}}$ of the pupil function $\mathcal{H}(\mathbf{u})$. The FFOCT image exhibits the same feature as an incoherent image but with a slightly better resolution. The FFOCT intensity PSF indeed scales as $|h|^4$ while the incoherent PSF scales as $|h|^2$. \alex{The particular case of a coma is investigated in details in Supplement 1. }

\subsection{Even phase pupil function}

On the contrary, for an even phase pupil function, the support of the ATF $\mathcal{H}_2(\mathbf{u})$ spans from 0 to \alex{2}. Unlike the odd case, aberrations do not \textit{a priori} filter the high spatial frequencies of the object. While the incoherent image is smoothed due to a loss of resolution, the coherent image is sharper but also noisy due to a bad recombination of the high spatial frequencies. Moreover, as the ATF is projected on the object's spectrum, the manifestation of aberrations will depend on \al{the} nature of the object. In the following, we consider the two asymptotic examples of fully coherent and incoherent objects. 

\subsubsection{Coherent object}
Let us first assume a coherent object displaying homogeneous correlation properties over its spatial frequency spectrum
\begin{equation}
    \mathcal{P}(\mathbf{u}-\mathbf{u'}/2)\mathcal{P}^*(\mathbf{u}+\mathbf{u'}/2)  = | \mathcal{P}(\mathbf{u}) |^2 \Gamma (\mathbf{u'}).
\end{equation}
$\Gamma(\mathbf{u'})$ is the normalized correlation function of $\mathcal{P}(\mathbf{u})$. Using that property, the FFOCT spectrum can be simplified as follows:
\begin{equation}
\mathcal{I}_F(\mathbf{u})=\Gamma (\mathbf{u}) \int d\mathbf{u'} \mathcal{H}_2(\mathbf{u'}-\mathbf{u}/2) \mathcal{H}_2^*(\mathbf{u'}+\mathbf{u}/2) |\mathcal{P}(\mathbf{u'})|^2.
\end{equation}
\al{The FFOCT image spectrum involves a correlation product of the pupil function $\mathcal{H}_2$ weighted by the intensity of the object spectrum $|\mathcal{P}(\mathbf{u})|^2$.} If the object is extended, its spatial frequency spectrum is of finite support around a direction $\mathbf{u_P}$ normal to its orientation. As in incoherent imaging, the aberrations manifest as a low pass filter but the cut-off $u_c$ will \al{correspond to} the local frequency correlation width in the vicinity of $\mathbf{u_P}$. Hence, if the object's spectrum is localized in a part of pupil plane where the ATF phase shift is minimal (or maximal), then its image will be less (more) impacted by aberrations than the corresponding incoherent image. \al{Note also that, for a point-like object ($|\mathcal{P}(\mathbf{u})|^2=1$), the FFOCT image becomes equivalent to an incoherent image with a confocal pupil function $\mathcal{H}_2$.}

\subsubsection{Random \al{phase} object}
Let us now consider an object of reflectivity $\rho(\mathbf{r})$ with a random phase $\phi(\mathbf{r})$:
\begin{equation}
    \rho(\mathbf{r})=|\rho(\mathbf{r})|e^{i\phi(\mathbf{r})},
\end{equation}
where $\phi(\mathbf{r})$ accounts for the random phase of the reflectivity: $ \left \langle e^{i\phi(\mathbf{r})} e^{-i\phi(\mathbf{r'})}\right \rangle=\delta(\mathbf{r}-\mathbf{r'})$. If we inject this expression of $\rho(\mathbf{r})$ into Eq.~\ref{terme_interference3}, the mean intensity of the FFOCT signal can be derived as follows:
\begin{equation}
\label{confinc}
    \langle I_F(\mathbf{r}_{\al{d}}) \rangle = I_0^2 \left [ |h_2|^2 * |\rho_m|^2 \right ](\mathbf{r}_{\al{d}}).
\end{equation}
\al{By analogy with Eq.~\ref{image_conventionnelle}, FFOCT image yields an incoherent-like image of the random phase object but with the confocal impulse response $h_2$.} This result is important since such a \al{random scattering} regime can be met in biological tissues made of a random distribution of unresolved scatterers. 

\subsection{General case}

\al{As seen above, for a random phase or point-like object, the FFOCT image displays an incoherent feature with an OTF equal to $\mathcal{H}_2 \otimes \mathcal{H}_2$ whatever the nature of aberrations.} On the contrary, for a coherent object, the phase parity of the pupil function is an important parameter to assess the impact of aberrations on the FFOCT image. However, there is not, in practice, such a thing as a purely anti-symmetric or symmetric aberration phase law. In a general case, the pupil function $\mathcal{H}$ can be decomposed as the sum of two pupil functions, $\mathcal{H}_+$ and $\mathcal{H}_-$, with symmetric and anti-symmetric phase distributions, respectively:
\begin{equation}
\mathcal{H}_{\pm}(\mathbf{u}) = [  \mathcal{H}(\mathbf{u}) \pm \mathcal{H}^*(-\mathbf{u})]/2 .
\end{equation}
Depending on the relative weight between each component, FFOCT will yield a coherent- or incoherent-like image of the object. To that aim, the ratio $\eta=||H_+||^2/||H_-||^2$ should be assessed, with $||H_\pm||^2>> \int d \mathbf{u} |\mathcal{H}_{\pm}|^2 $.  If $\rho<<1$, the FFOCT image will be equivalent to a confocal incoherent image with an intensity PSF scaling as $|h|^4$ (see Sec.~\ref{odd}).  If $\rev{\eta}>>1$, the FFOCT image displays a coherent-like feature for a coherent object. A strong gain in resolution can be expected if the object's spectrum lies in a pupil area where $\mathcal{H}(\mathbf{u})$ exhibits a minimal phase shift. If $\rev{\eta} \sim 1$, the resulting image will be the superimposition of incoherent and coherent confocal images. \alex{This particular case is investigated in details in Supplement 1. }

\section{High-order aberrations}
\begin{figure}[htbp]
\centering
\includegraphics[width=0.5\linewidth]{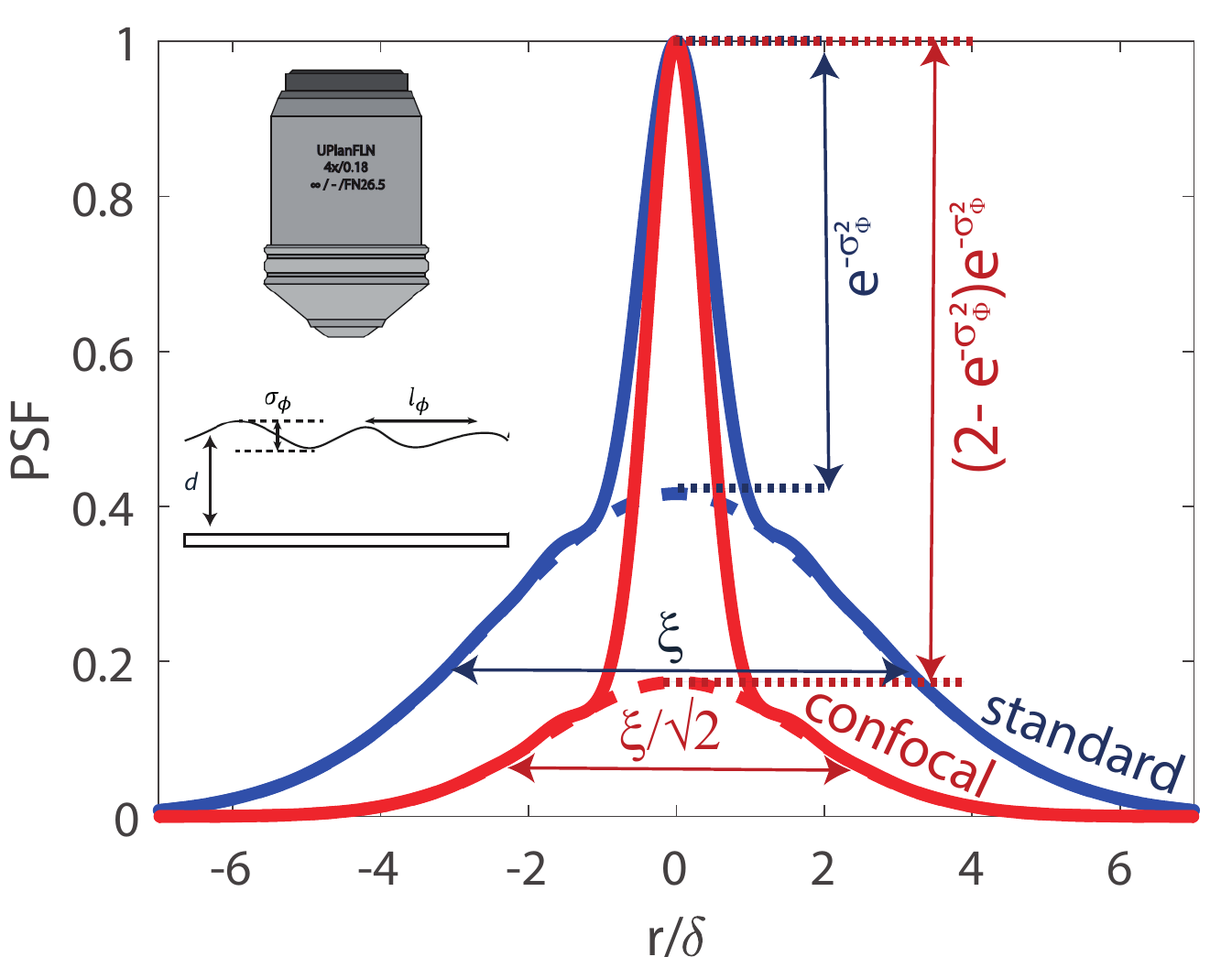}
\caption{Incoherent and FFOCT PSFs induced by a thin aberrating layer. The parameters of the random phase screen are the following: $\sigma_\phi=0.5$ and  $u_\phi=0.5$. The incoherent intensity PSF $|h|^2$ (blue curve) is compared to the FFOCT PSF $|h|^4$ (red curve). The aberrated components of each PSF, $|h_A|^2$ and $|h_A|^4$, are superimposed onto the corresponding PSFs as a dashed line.}
\label{fig5}
\end{figure}
In view of deep imaging applications in biological tissues, high-order aberrations are now investigated. To that aim, the basis of Zernike polynomials is no longer adequate and a random Gaussian model is more realistic. The building block for such a model is the well-known random phase screen~\cite{Schott2015}. It models a thin aberrating layer that we suppose located at a distance $d$ from the focal plane \alex{(see the inset of Fig.~\ref{fig5})}. The latter one can be modelled as a thin phase screen of transmittance
\begin{equation}
\label{thin}
   \mathcal{H}(\mathbf{r}_\phi)=\exp\left [ i\phi(\mathbf{r}_\phi \rev{)} \right ].
\end{equation}
where $\phi(\mathbf{r}_\phi)$ is a local (real) phase shift in the aberrating layer plane. In the following, $\phi(\mathbf{r}_\phi)$ is assumed to follow a Gaussian random statistics of zero mean and variance $\sigma_\phi^2$ with a characteristic coherence length $l_\phi$. If the aberrating layer is sufficiently far from the focal plane ($d>>\lambda$), geometrical optics can be used to rescale the transmittance $\mathcal{H}(\mathbf{r}_\phi)$ into the pupil plane, \rev{such} that:
\begin{equation}
\label{thin2}
   \mathcal{H}(\mathbf{u}_\phi)=\exp\left [ i\phi( \mathbf{u} /d) \right ].
\end{equation}
The phase of the pupil function still follows a Gaussian statistics with the same variance $\sigma_\phi^2$ and a coherence length $u_\phi=l_\phi/d$. The incoherent OTF, $\mathcal{H} \otimes \mathcal{H}$, is given by~\cite{mertz2015field}
\begin{equation}
\label{OTFthin0}
 \mathcal{H} \otimes \mathcal{H}(\mathbf{u}) = e^{-\sigma_\phi^2 [ 1-\gamma_0(\mathbf{u})]} ,
\end{equation}
where $\gamma_0(\mathbf{u})=\exp(-u^2/u_\phi^2)$ is the normalized auto-correlation function of the phase $\phi$ in the pupil plane. To gain further insight into the physics, Eq.~\ref{OTFthin} can be recast as follows~\cite{mertz2015field}
\begin{equation}
\label{OTFthin}
\mathcal{H} \otimes \mathcal{H} (\mathbf{u}) \simeq e^{-\sigma_\phi^2} + (1-e^{-\sigma_\phi^2} ) e^{-u^2/u_c^2},
\end{equation}
with
\begin{equation}
    u_c=\frac{u_\phi}{(1+\sigma_\phi^2)},
\end{equation} 
the coherence length of the pupil function. Equation~\ref{OTFthin} shows that the reflected wave-field is made of ballistic (left term) and scattered (right term) components. The ballistic component is attenuated by a factor $e^{-\sigma_\phi^2}$ because of scattering while the rest of the wave-field is scattered. 

The incoherent PSF $|h|^2$ can be deduced from the Fourier transform of the OTF. It combines an attenuated version of the diffraction-limited ballistic PSF, $|h_0|^2$ , on top of a wider pedestal (see Fig.~\ref{fig5}) resulting from the aberration induced by the phase screen:
\begin{equation}
\label{PSFthin}
    |h(\mathbf{r})|^2 \simeq e^{-\sigma_\phi^2} |h_0(\mathbf{r})|^2 + (1-e^{-\sigma_\phi^2}) \underbrace{\frac{\exp \left ( - r^2/ \xi^2 \right ) }{\pi \xi^2}}_{=|h_A(\mathbf{r})|^2},
\end{equation}
where $\xi=\lambda / (\pi u_c)$ is the spatial extension of the aberrated incoherent PSF $|h_A|^2$. Figure~\ref{fig5} shows the typical superimposition of the ballistic and aberrated components in an incoherent PSF resulting from the high-order phase distortions induced by a thin aberrating layer. 

For a fully incoherent or point-like object, the FFOCT PSF is the square of the incoherent PSF: $|h_2|^2= |h|^4$ (Eq.~\ref{confinc}). Using Eq.~\ref{PSFthin}, the following expression for $|h_2|^2$ can be derived:
\begin{eqnarray}
    |h_2(\mathbf{r})|^2 & = & e^{-2\sigma_\phi^2} |h_0(\mathbf{r})|^4 \nonumber \\
    & + & 2 e^{-\sigma_\phi^2} (1-e^{-\sigma_\phi^2}) \frac{\exp \left ( -r^2/ \xi^2 \right ) }{\pi \xi^2} |h_0(\mathbf{r})|^2 \nonumber \\
    & + & \left (1-e^{-\sigma_\phi^2} \right )^2 \frac{\exp \left ( -2r^2/ \xi^2 \right ) }{\pi^2 \xi^4} .
    \label{last}
\end{eqnarray}
The first term  accounts for the ballistic light whose attenuation $ e^{-2\sigma_\phi^2}$ account for the travel back and forth through the aberration layer of the coherent wave. As demonstrated above, it exhibits a thinner intensity PSF $|h_0|^4$ than the incoherent standard PSF $|h_0|^2$. The second term combines an incident ballistic path and a reflected scattered path \rev{as well as the reciprocal trajectory (scattered incident path and ballistic reflected path)}. The associated PSF is the product of the ballistic and aberrated PSFs. $|h_0|^2$ being much thinner than $|h_A|^2$, its width is thus close to the diffraction-limited PSF $|h_0|^2$.  At last, the third term is due to scattering and corresponds to the confocal aberrated PSF $|h_A|^4$. The diffraction-limited component combines the two first terms and scales as $e^{-\sigma_\phi^2}(2-e^{-\sigma_\phi^2})$. The confocal filter induced by the spatial incoherence of the incident wave-field in FFOCT allows to increase the weight of the diffraction-limited PSF by a factor $\left (2-e^{-\sigma_\phi^2} \right)$ compared to standard incoherent imaging. Moreover, the characteristic width of the aberrated PSF is decreased by a factor two. The impact of the aberrating layer is thus drastically reduced in FFOCT. Figure~\ref{fig5} illustrates this fact by showing an example of FFOCT PSF, $|h|^4$ and compare it with the orginal incoherent PSF $|h|^2$. A clear improvement is found both in terms of resolution (reduction of the PSF extension) and contrast (increase of the weight of ballistic light).

\alex{Note that a similar analytical expression of the FFOCT signal has been derived in three-dimensional scattering media by Andersen \textit{et al.}~\cite{Andersen}. It basically consists in replacing the ballistic attenuation term $e^{-\sigma_\phi^2}$ by $e^{-L/ \ell_s}$ in Eq.~\ref{last} with $L$, the depth between the scattering medium surface and the coherence plane and $\ell_s$, the scattering mean free path. Equation~\ref{last} thus holds in scattering media. The effect of scattering for deep imaging of biological tissues can actually be grasped by stacking random phase screens~\cite{Schott2015}.} 

\section{Discussion}

\alexx{This paper proposes a Fourier optics model to account for the impact of phase aberrations on reflection optical imaging methods with a particular focus on FFOCT. To that aim, two kind of objects have been considered: A spoke target and isolated point-like scatterers. 
The former object approximates a radially varying measure of spatial frequency contrast in the presence of noise and system aberrations. For incoherent imaging, aberrations behave as a spatial frequency filter. The image of the spoke target thus directly provides the corresponding cutoff and highlights potential contrast inversions~\cite{Goodman}. For coherent imaging, the impact of phase aberrations is more tricky. They do not filter the spatial frequency components of the object but induces a phase shift between them. Contrast inversion in the intensity image is an indirect manifestation of phase distortions in the ATF but, as shown in this paper, this phenomenon is object-dependent. }

\alexx{A point-like scatterer is a more relevant observable to investigate the impact of phase aberrations in coherent imaging. Indeed, the spatial frequency components shall add constructively at a single point in the image. The least phase shift between them has a direct consequence on the image amplitude of the point-like scatterer. However, one has to be careful to not only consider the width of the main lobe in this imaging PSF but also quantify the occurrence and level of secondary lobes. For some peculiar aberration distribution~\cite{Xiao_16_osa}, a diffraction-limited resolution seems to be reached but this is at the cost of strong secondary lobes and of a bad constrast. The two experimental configurations, spoke and point-like target, are thus of interest and complementary. }

\alexx{In FFOCT and more generally, in coherent imaging, one has to be careful with respect to claims about the robustness to aberrations. For instance, the random distribution of scatterers in biological media gives rise to an image of speckle whose rich spatial frequency content is not altered by phase aberrations. By no means, the impact of aberrations can thus be assessed by looking at the speckle grain size. Aberrations only result in a bad recombination of each spatial frequency component induced by phase distortions between them. Only the imaging PSF provides a direct measure of this bad recombination. The latter one can then be quantified by parameters such as the Strehl ratio~\cite{mahajan1982strehl}. Note that, in a reflection configuration, the imaging PSF can be retrieved, not only in presence of isolated point-like scatterers, but also for specular reflectors~\cite{Badon2019} or in random scattering media~\cite{Lambert2020}. A virtual point-like scatterer can actually be synthesized at any point of the inspected medium by means of the distortion matrix concept. A map of the local Strehl ratio is built and quantifies locally the level of aberrations~\cite{Lambert2020}.}

\alexx{Although the Fourier optics model presented in this paper is of interest to assess the impact of aberrations in reflection microscopy, it does not predict the performance of FFOCT in biological media. First, only planar objects have been here considered. Yet FFOCT imaging in three-dimensional scattering media has been recently tackled~\cite{Tricoli2019,zhou2020unified}. Interestingly, the theoretical expression of the FFOCT image derived in Eq.~\ref{terme_interference3} holds in such media. Only the ballistic attenuation of the incident and reflected light across the medium has to be included. However, as in the current paper, the deformation of the coherence plane induced by aberrations and/or scattering events ahead of the focal plane (Fig.~\ref{fig3}) is not taken into account. \alex{Regarding that issue, the incorporation of adaptive optics tools in a FFOCT system is of particular interest~\cite{mece2020coherence,Scholler:20}. Besides aberration effects,} multiple scattering phenomena should be also included in the modeling as it constitutes a fundamental limit for deep optical imaging. In that perspective, Badon \textit{et al.}~\cite{badon2017multiple} recently predicted the evolution in depth of the single-to-multiple scattering ratio for different reflection imaging techniques including FFOCT. \alex{To overcome multiple scattering in FFOCT, a matrix imaging strategy can be a relevant tool~\cite{kang2015imaging,Badon_svd_16,Badon2019}.} At last, the signal-to-noise ratio in OCT is also an important issue to predict the performance of optical imaging techniques in the quest for deep tissue imaging. With respect to that issue, the Fourier domain FFOCT presents a clear advantage over its time domain counterpart~\cite{deBoer_03}. In the former case, the signal-to-noise ratio is actually shown to increase linearly with the number of independent wavelengths over which the FFOCT signal is recorded~\cite{Izatt2008}.}

\section{Conclusion}
\alexx{In this paper, we report on the manifestation of aberrations in FFOCT by means of three experiments, one considering a spoke target and, the others, isolated point-like scatterers as the object to image. While FFOCT seems particularly robust to low-order symmetric aberrations (such as defocus) on a coherent object such as a resolution target, the FFOCT image of a point-like scatterer tends to be equivalent to its incoherent counterpart for a large defocus. These experimental results are interpreted by means of a simple Fourier optics model that allows to compare FFOCT with other standard reflection imaging techniques. \rev{Under a spatially-incoherent illumination,} FFOCT is shown to be equivalent to a time-gated confocal microscope, which results in the extension by a factor two of the OTF and ATF supports compared to \rev{conventional microscopy and coherent FFOCT}, respectively. This leads to a gain in resolution by almost a factor two compared to \rev{these} standard imaging techniques. While aberrations manifest themselves as a spatial frequency filter in incoherent imaging, aberrations generally induce a phase distortion between the different spatial frequency components of the object in FFOCT. For a coherent object such as a spoke target, aberrations do not alter drastically the intensity image of the target. On the contrary, for a point-like scatterer, these phase distortions induce a bad recombination between the spatial frequencies at the focusing point. This study proposes a rigorous framework to understand the impact of aberrations in FFOCT and, more generally, in reflection optical imaging. The developed model will be an important tool for the design of efficient adaptive optics schemes in FFOCT or novel matrix methods for computational imaging.}

\begin{acknowledgments}
The authors wish to thank Serge Meimon \al{and Paul Balondrade} for fruitful discussions.
The authors are grateful for the funding provided by Labex WIFI (Laboratory of Excellence within the French Program Investments for the Future) (ANR-10-LABX-24 and ANR-10-IDEX-0001-02 PSL*). This project has received funding from the European Research Council (ERC) under the European Union's Horizon 2020 research and innovation programme (grant agreements No. 610110 and No. 819261). 

\end{acknowledgments}

\clearpage 

\clearpage

\renewcommand{\thetable}{S\arabic{table}}
\renewcommand{\thefigure}{S\arabic{figure}}
\renewcommand{\theequation}{S\arabic{equation}}
\renewcommand{\thesection}{S\arabic{section}}

\setcounter{equation}{0}
\setcounter{figure}{0}

\begin{center}
\Large{\bf{Supplementary Information}}
\end{center}

This document provides supplementary information on: (\textit{i}) the experimental set up used for the spoke target experiment; (\textit{ii}) the comparison between theory and experiment for resolution target images under astigmatism and coma; (\textit{iii}) the experimental configuration used for the gold nano-bead experiment and its theoretical modelling.

\section{Spoke target experiment}

\begin{figure}[htbp]
  \centering
  \includegraphics[width=0.5\linewidth]{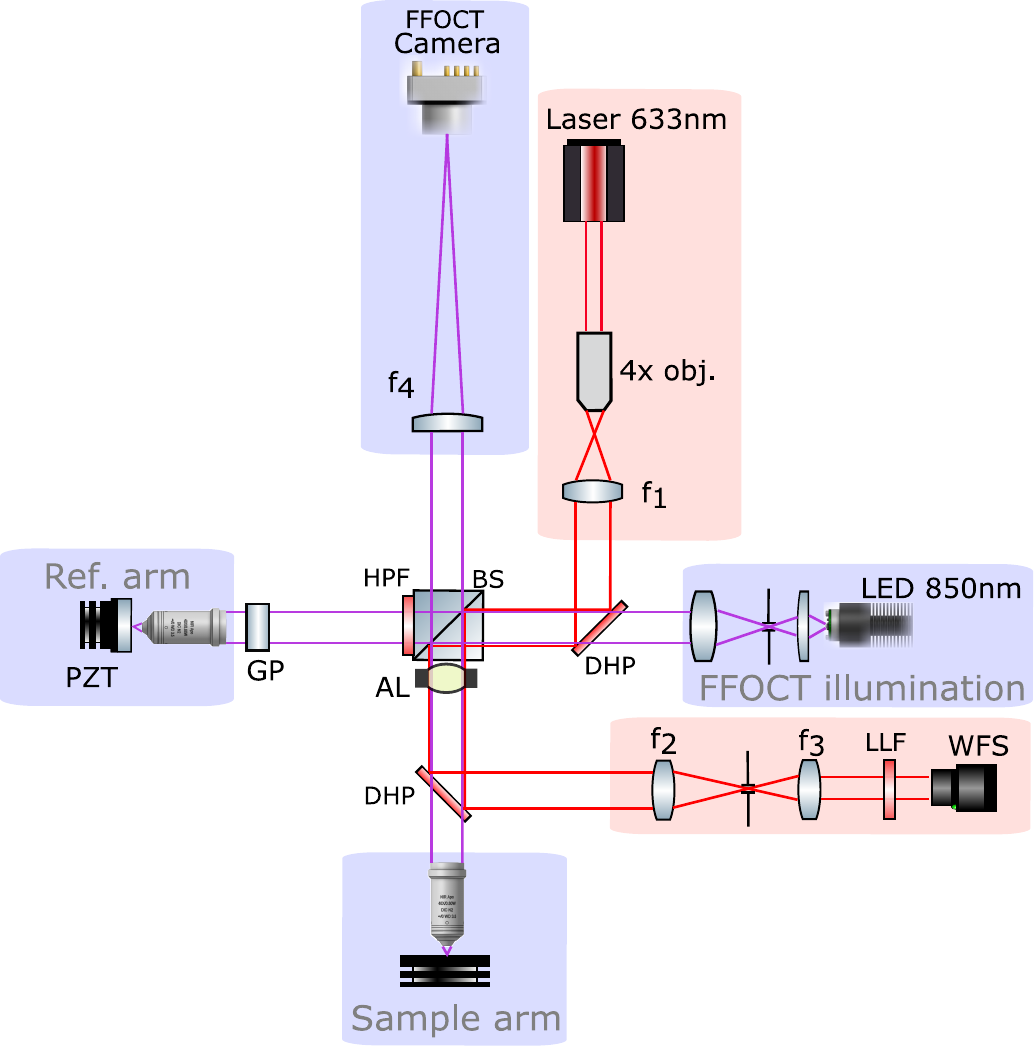}
    \caption{Experimental setup \alex{combining FFOCT and AO. Blue and red parts correspond to FFOCT and AO arms, respectively.} WFS: wavefront sensor - AL: adaptive lens - GP: glass plate used for dispersion compensation - DHP: dichroic high pass mirror - HPF: high pass filter - BS: beam splitter - LLF: laser line filter - $f_1 = 150$ mm - $f_2 = 150$ mm - $f_3 = 60$ mm - $f_4 = 350$ mm.}
  \label{fig:AO_FFOCT}
\end{figure}

\alex{The spoke target experiment combines a standard FFOCT setup with an adaptive optics (AO) system.} To do so, a Shack-Hartmann wavefront sensor (SH-WFS) and an adaptive lens (AL) (both from Dynamic Optics srl, Italy) have been used~\cite{Bonora_15, Verstraete_17}. The setup is depicted in Fig.~\ref{fig:AO_FFOCT}. The FFOCT part shown in blue consists of a Linnik interferometer (with Olympus UCPLFLN20X objectives) illuminated by a LED source (\textit{M850L3}, Thorlabs). The sample arm is mounted on a translation stage (T-LSR150B, Zaber Technologies) and the reference mirror is mounted on a piezoelectric translation stage (STr-25, Piezomechanik) for phase-shifting. 

The AO part is shown in red. The beam from a He:Ne laser ($633$ nm) is expanded and collimated using a 4$\times$ microscope objective and a $f_1=100$ mm lens and used as a plane wave reference for the AO part. This beam is combined with the OCT illumination using a dichroic mirror (\textit{FM02R}, Thorlabs) before the 45:55 beam splitter (\textit{BP145B2}, Thorlabs) and blocked from entering the reference arm using a high pass filter as we wish to image the aberrated PSF in the sample arm only. The AL is set a few centimeters back from the pupil plane of the sample arm objective. The dispersion is compensated by a $2$-mm-thick glass plate in the reference arm. In the sample arm, the beam is splitted by the same dichroic mirror as in the illumination arm. The pupil plane is relayed using a $4f$ telescope with a diaphragm (to filter out parasite reflections) onto the SH-WFS lenslet array. 
\noindent The remaining LED infrared photons are filtered out using a laser line filter centered at 633 nm. In this configuration, we do not block the laser photons from impinging the OCT camera so we can image the sample arm PSF and monitor it through our FFOCT acquisition software \cite{Scholler_FFOCT_Zenodo}. This setup allows to jointly acquire the FFOCT image and the associated pupil function. Note that the SH-WFS arm could be put at the exit of the interferometer to perform AO correction but the measured wavefront would correspond to the round trip and could therefore not be used in this study to extract the pupil function \alex{$\mathcal{H}$}.

The AL was calibrated following the procedure described in \cite{Bonora_15}. The wave-front is controlled in closed loop after the acquisition of the influence functions of each single actuator. \alexx{The aberrations are mapped onto the 1144 centroids of the SH-WFS, which corresponds to a pupil diameter made of 38 centroids. The recorded aberrations are then projected on the \all{18} first (ANSI/OSA index) Zernike polynomials~\cite{noll1976zernike}}. In its rest position (i.e. when all the applied voltage are 0), the AL exhibits \alex{phase} aberrations of 0.65 root mean square (RMS) with mainly astigmatism \all{($Z^2_2$, $Z^{-2}_2$)} and defocus ($Z^0_2$). The first step for operating the system is to correct for the AL initial aberrations. Correcting these aberrations leads to a wavefront distorsion of 0.05 $\lambda$ RMS. It requires 55\% RMS of the lens dynamic thus leaving room for introducing deterministic wavefront errors.  The pupil function \alex{$\mathcal{H}$} is then constructed without the piston as the latter one cannot be measured with a SH-WFS. 

The incoherent image of the spoke target (Fig.~\ref{fig2}c of the accompanying paper) is recorded by blocking the reference arm. The FFOCT image is captured with a linear 4 phase buckets scheme in order to extract the amplitude term \cite{CREATH_1988}. Using the experimental setup of Fig.~\ref{fig:AO_FFOCT}, it is therefore possible to jointly acquire standard incoherent and FFOCT images along with the associated pupil function $\mathcal{H}$.

\section{Spoke target imaging under astigmatism and defocus}

\begin{figure*}[htbp]
  \centering
  \includegraphics[width=\linewidth]{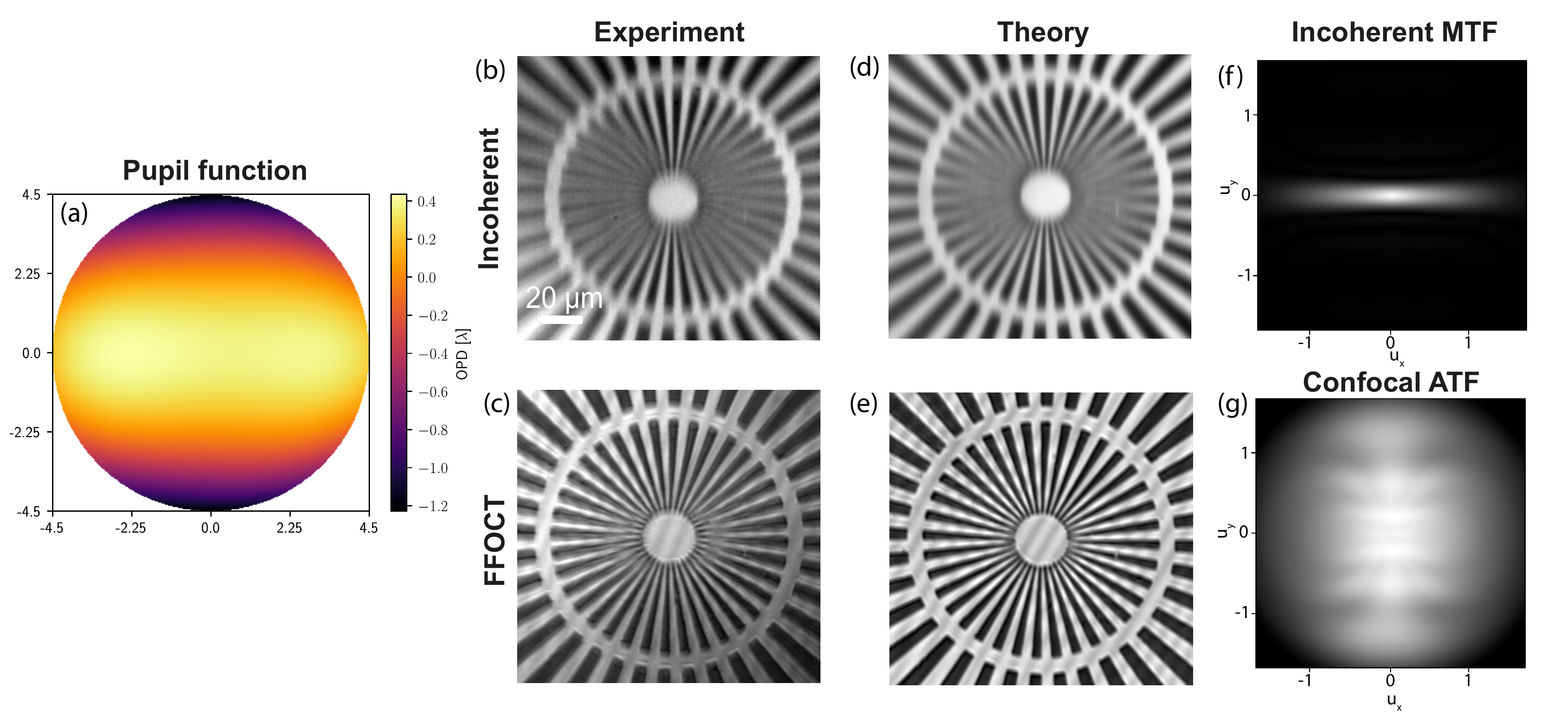}
    \caption{Imaging of the spoke target in presence of defocus and astigmatism. (a) Pupil function. (b,c) Experimental incoherent and FFOCT images of the target. (d,e) Theoretical ioncoherent and FFOCT images. (f) Modulation transfer function ($[\mathcal{H} \otimes \mathcal{H}] (\mathbf{u})$). (g) Confocal ATF ($\mathcal{H}_2(\mathbf{u})=[\mathcal{H} * \mathcal{H}] (\mathbf{u})$).}
  \label{figS2}
\end{figure*}

Figure~\ref{figS2} shows the experimental and theoretical results when the AL induces the combination of a vertical astigmatism  \all{$Z^{2}_2$} ($0.3\lambda$) and a defocus $Z^0_2$ ($-0.7\lambda$). The corresponding pupil function $\mathcal{H}(\mathbf{u})$ is displayed in Fig.~\ref{figS2}a. The experimental incoherent and FFOCT images are displayed in Fig.~\ref{figS2}b and c, respectively. These experimental results are in good agreement with the theoretical images computed from the Fourier optics model described in the accompanying paper (Fig.~\ref{figS2}d and e). Not surprisingly, the impact of aberrations on the incoherent image is much stronger in the vertical direction because of the astigmatism orientation. As displayed by the theoretical incoherent MTF ($|\mathcal{H} \otimes \mathcal{H}|$) displayed in Fig.~\ref{figS2}f, the astigmatism acts as an anisotropic spatial frequency filter with a much lower frequency cutoff $u_{c}$ in the y-direction. On the contrary, FFOCT is very robust to the symmetric aberration phase law (Fig.~\ref{figS2}d and e). The corresponding ATF, $\mathcal{H}_2=\mathcal{H} \otimes \mathcal{H}$, extends from $u=0$ to the confocal cut-off $u=2$. Most of the object's frequency spectrum is thus preserved by the FFOCT system and the intensity image of the target is a reliable estimator of its reflectivity.

\section{Spoke target imaging under coma and defocus}

\begin{figure*}[htbp]
  \centering
  \includegraphics[width=\linewidth]{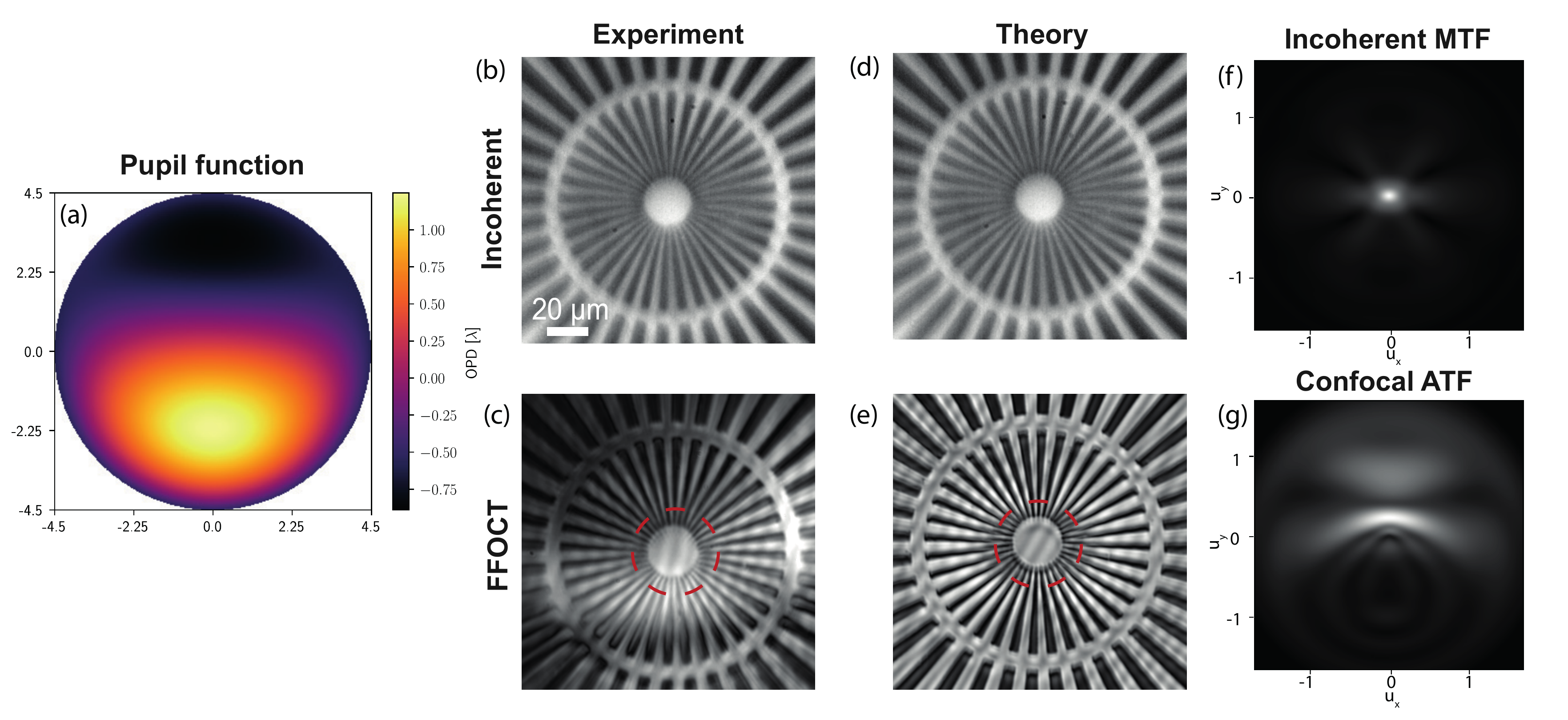}
    \caption{Imaging of the spoke target in presence of defocus and coma. (a) Pupil function. (b,c) Experimental incoherent and FFOCT images of the target. (d,e) Theoretical incoherent and FFOCT images.(f) Modulation transfer function ($[\mathcal{H} \otimes \mathcal{H}] (\mathbf{u})$). (g) Confocal ATF ($\mathcal{H}_2(\mathbf{u})=[\mathcal{H} * \mathcal{H}] (\mathbf{u})$).}
  \label{figS3}
\end{figure*}

Figure~\ref{figS3} shows the experimental and theoretical results when the AL induces the combination of a vertical coma $Z^{1}_3$ \alexx{($0.3$ $\lambda$) and a defocus $Z_2^0$ ($-0.7\lambda$)}. The corresponding pupil function $\mathcal{H}(\mathbf{u})$ is displayed in Fig.~\ref{figS3}a.  The experimental and theoretical incoherent images are shown in Fig.~\ref{figS3}b and d. Both show an excellent agreement and highlight the important loss of resolution induced by the coma. The corresponding MTF, displayed in Fig.~\ref{fig3}f, shows the low-pass spatial frequency filtering operated by the coma with a nearly isotropic spatial frequency cutoff $u_c=0.1$. 

With regards to FFOCT, the agreement between the experimental result (Fig.~\ref{figS3}c) and the theoretical prediction (Fig.~\ref{figS3}e) is less obvious. \all{Unlike the theoretical prediction, the experimental image actually shows contrast fluctuations across the field-of view. One possible reason} is that the AL is not exactly placed in the pupil plane of the microscope. The associated PSF is thus not fully spatially invariant, hence a contrast fluctuation across the field-of-view. Another explanation is the deformation of the coherence plane induced by the coma that is not taken into account by our model.

While the theoretical study in the accompanying paper shows that an odd aberration phase law like coma should act similarly for incoherent and coherent imaging, Fig.~\ref{figS3} highlights a strong difference between the incoherent and FFOCT images. The reason is the presence of a defocus on top of the coma in the pupil function (Fig.~\ref{figS3}a). While, for a pure coma, the incoherent MTF and confocal ATF should be identical, the comparison between Figs.~\ref{figS3}f and g shows that this is not at all the case here. Consequently, albeit modulated, the confocal ATF spreads over a much wider support than the incoherent MTF. This explains the much better quality of the FFOCT image compared to the incoherent one. Nevertheless, even in presence of defocus, the coma should give rise, in theory, to a loss of contrast at high spatial frequencies (see the finest details of the target contained in the red dashed circle in Fig.~\ref{figS3}e). This effect is less obvious experimentally (Fig.~\ref{figS3}e) since it is probably hidden by the aforementioned intensity fluctuations across the field-of-view.

\section{Gold nanobead experiment}

\begin{figure*}[htbp]
  \centering
  \includegraphics[width=0.5\linewidth]{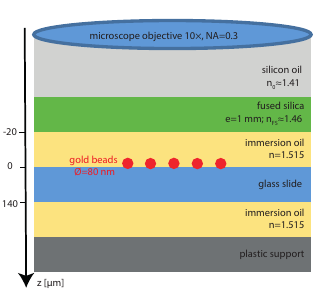}
    \caption{\all{Experimental configuration of the FFOCT  gold nanobead experiment.}}
  \label{figJM0}
\end{figure*}

\all{Similarly to the experiment performed in Ref.~\cite{Xiao_16_osa}, a commercial LLtech FFOCT system Light-CT Scanner~\cite{LLTech} ($\lambda \sim 626$ nm, $\Delta \lambda \sim 12$ nm) is used to image 80 nm-diameter gold nanoparticles under different level of defocus. The gold nanoparticle solution has been first diluted and dried on a coverslip so that an isolated nano-particle can be imaged by the FFOCT system through an immersion microscope objective ($\sin \alpha=0.3$). The corresponding experimental configuration is described in Fig.~\ref{figJM0}. By moving the sample stage, several amounts of defocus (from $z=0$ to 70 $\mu$m) have been applied to the targeted particle. The length of the reference arm is also shifted by a distance $nz$ in order to match the coherence plane and the bead position, with $n=1.515$ the optical index of the immersion oil. The FFOCT image of the gold nanobead is displayed in Fig.~\ref{figJM} of the accompanying paper for different values of defocus.}

\end{document}